%%%%%%%%%%%%%%%%%
%% This file runs in plain Tex and requires no input macros.
%% It already includes a shortened version of jnl.tex and
%% reforder.tex
%%%%%%%%%%%%%%%%%%%%%%%%%%%%%%%%%%%%%%%%%%%%%%%%%%%%%%%%%%%%%%
\font\twelverm=cmr10 scaled 1200    \font\twelvei=cmmi10 scaled 1200
\font\twelvesy=cmsy10 scaled 1200   \font\twelveex=cmex10 scaled 1200
\font\twelvebf=cmbx10 scaled 1200   \font\twelvesl=cmsl10 scaled 1200
\font\twelvett=cmtt10 scaled 1200   \font\twelveit=cmti10 scaled 1200
\font\twelvesc=cmcsc10 scaled 1200  %\font\twelvesf=cmssmc10 scaled 1200
\skewchar\twelvei='177   \skewchar\twelvesy='60
     
%  Define \...point macros to change fonts and spacings consistently
     
\def\twelvepoint{\normalbaselineskip=12.4pt plus 0.1pt minus 0.1pt
  \abovedisplayskip 12.4pt plus 3pt minus 9pt
  \belowdisplayskip 12.4pt plus 3pt minus 9pt
  \abovedisplayshortskip 0pt plus 3pt
  \belowdisplayshortskip 7.2pt plus 3pt minus 4pt
  \smallskipamount=3.6pt plus1.2pt minus1.2pt
  \medskipamount=7.2pt plus2.4pt minus2.4pt
  \bigskipamount=14.4pt plus4.8pt minus4.8pt
  \def\rm{\fam0\twelverm}          \def\it{\fam\itfam\twelveit}%
  \def\sl{\fam\slfam\twelvesl}     \def\bf{\fam\bffam\twelvebf}%
  \def\mit{\fam 1}                 \def\cal{\fam 2}%
  \def\sc{\twelvesc}               \def\tt{\twelvett}
  \def\sf{\twelvesf}
  \textfont0=\twelverm   \scriptfont0=\tenrm   \scriptscriptfont0=\sevenrm
  \textfont1=\twelvei    \scriptfont1=\teni    \scriptscriptfont1=\seveni
  \textfont2=\twelvesy   \scriptfont2=\tensy   \scriptscriptfont2=\sevensy
  \textfont3=\twelveex   \scriptfont3=\twelveex  \scriptscriptfont3=\twelveex
  \textfont\itfam=\twelveit
  \textfont\slfam=\twelvesl
  \textfont\bffam=\twelvebf \scriptfont\bffam=\tenbf
  \scriptscriptfont\bffam=\sevenbf
  \normalbaselines\rm}
     
%       tenpoint

%%
%%      Various internal macros
%%
     
\def\beginlinemode{\endmode
  \begingroup\parskip=0pt \obeylines\def\\{\par}\def\endmode{\par\endgroup}}
\def\beginparmode{\endmode
  \begingroup \def\endmode{\par\endgroup}}
\let\endmode=\par
{\obeylines\gdef\
{}}
\def\singlespace{\baselineskip=\normalbaselineskip}

\def\oneandahalfspace{\baselineskip=\normalbaselineskip
  \multiply\baselineskip by 3 \divide\baselineskip by 2}
\def\doublespace{\baselineskip=\normalbaselineskip \multiply\baselineskip by 2}

\newcount\firstpageno
\firstpageno=2
\footline={\ifnum\pageno<\firstpageno{\hfil}\else{\hfil\twelverm\folio\hfil}\fi}
\def\toppageno{\global\footline={\hfil}\global\headline
  ={\ifnum\pageno<\firstpageno{\hfil}\else{\hfil\twelverm\folio\hfil}\fi}}
\let\rawfootnote=\footnote              % We must set the footnote style
\def\footnote#1#2{{\rm\singlespace\parindent=0pt\parskip=0pt
  \rawfootnote{#1}{#2\hfill\vrule height 0pt depth 6pt width 0pt}}}
\def\raggedcenter{\leftskip=4em plus 12em \rightskip=\leftskip
  \parindent=0pt \parfillskip=0pt \spaceskip=.3333em \xspaceskip=.5em
  \pretolerance=9999 \tolerance=9999
  \hyphenpenalty=9999 \exhyphenpenalty=9999 }
\def\dateline{\rightline{\ifcase\month\or
  January\or February\or March\or April\or May\or June\or
  July\or August\or September\or October\or November\or December\fi
  \space\number\year}}
\def\received{\vskip 3pt plus 0.2fill
 \centerline{\sl (Received\space\ifcase\month\or
  January\or February\or March\or April\or May\or June\or
  July\or August\or September\or October\or November\or December\fi
  \qquad, \number\year)}}
     
%%
%%      Page layout, margins, font and spacing (feel free to change)
%%
     
\hsize=6.5truein
%\hoffset=1truein
\vsize=8.5truein  
%\voffset=-1.0truein
\parskip=\medskipamount
\def\\{\cr}
\twelvepoint            % selects twelvepoint fonts (cf. \tenpoint)
\doublespace            % selects double spacing for main part of paper (cf.
                        %       \singlespace, \oneandahalfspace)
\overfullrule=0pt       % delete the nasty little black boxes for overfull box

\def\title                      %  Title on title page
  {\null\vskip 3pt plus 0.2fill
   \beginlinemode \doublespace \raggedcenter \bf}
     
\def\author                     %  Author(s) name(s)  on title page
  {\vskip 3pt plus 0.2fill \beginlinemode
   \singlespace \raggedcenter\sc}
     
\def\affil                      % Affiliations (can intermix with \author)
  {\vskip 3pt plus 0.1fill \beginlinemode
   \oneandahalfspace \raggedcenter \sl}
     
\def\abstract                   % Begin abstract
  {\vskip 3pt plus 0.3fill \beginparmode
   \singlespace ABSTRACT: }
     
\def\endtopmatter               % End title page, begin body of paper
  {\endpage                     %       This subsumes \body
   \body}
     
\def\body                       % Begin text body;  can be used to end
  {\beginparmode}               % \title, \author, \affil, \abstract,
                                % \reference, or \figurecaption modes
     
\def\head#1{                    % Head;  NOTE enclose the text in {}
  \goodbreak\vskip 0.5truein    %  e.g., \head{I. Introduction}
  {\immediate\write16{#1}
   \raggedcenter \uppercase{#1}\par}
   \nobreak\vskip 0.25truein\nobreak}
     
\def\subhead#1{                 % Subhead;  NOTE enclose the text in {}
  \vskip 0.25truein             % e.g., \subhead{A. History of the Problem}
  {\raggedcenter {#1} \par}
   \nobreak\vskip 0.25truein\nobreak}
     
\def\beginitems{
\par\medskip\bgroup\def\i##1 {\item{##1}}\def\ii##1 {\itemitem{##1}}
\leftskip=36pt\parskip=0pt}
\def\enditems{\par\egroup}
     
\def\beneathrel#1\under#2{\mathrel{\mathop{#2}\limits_{#1}}}
     
\def\refto#1{$^{#1}$}           % For references in text as superscript
     
\def\references                 % Begin references -- basic format is Phys Rev
  {\head{References}            % I.e., volume, page, year (space after commas).
   \beginparmode
   \frenchspacing \parindent=0pt \leftskip=1truecm
   \parskip=8pt plus 3pt \everypar{\hangindent=\parindent}}

\gdef\refis#1{\item{#1.\ }}                     % Ref list numbers.
     
\gdef\journal#1, #2, #3, 1#4#5#6{               % Journal reference.  Comma sets
    {\sl #1~}{\bf #2}, #3 (1#4#5#6)}            % off: name, vol, page, year

\gdef\refa#1, #2, #3, #4, 1#5#6#7.{\noindent#1, #2 {\bf #3}, #4 (1#5#6#7).\rm} 
%refa: type in: name, 
%journal, vol, page, year
%prints out in same order

\gdef\refb#1, #2, #3, #4, 1#5#6#7.{\noindent#1 (1#5#6#7), #2 {\bf #3}, #4.\rm} 
%refb: reads in same
%prints out name (year) etc.

\def\pr{\journal Phys.Rev., }

\def\jmp{\journal J.Math.Phys., }

\def\endreferences{\body}

\def\endpage                    %  Eject a page
  {\vfill\eject}
     
\def\endpaper                   %  Ways to say goodbye
  {\endmode\vfill\supereject}

\def\ref#1{Ref.~#1}                     %       for inline references
\def\Ref#1{Ref.~#1}                     %       ditto
\def\[#1]{[\cite{#1}]}
\def\cite#1{{#1}}
\def\(#1){(\call{#1})}
\def\call#1{{#1}}
\def\taghead#1{}
\def\frac#1#2{{#1 \over #2}}
\def\half{{\frac 12}}

\def\12{{1\over2}}

\catcode`@=11
\newcount\r@fcount \r@fcount=0
\newcount\r@fcurr
\immediate\newwrite\reffile
\newif\ifr@ffile\r@ffilefalse
\def\w@rnwrite#1{\ifr@ffile\immediate\write\reffile{#1}\fi\message{#1}}

\def\writer@f#1>>{}
\def\referencefile{%			  Stuff to write .REF file
  \r@ffiletrue\immediate\openout\reffile=\jobname.ref%
  \def\writer@f##1>>{\ifr@ffile\immediate\write\reffile%
    {\noexpand\refis{##1} = \csname r@fnum##1\endcsname = %
     \expandafter\expandafter\expandafter\strip@t\expandafter%
     \meaning\csname r@ftext\csname r@fnum##1\endcsname\endcsname}\fi}%
  \def\strip@t##1>>{}}

\def\citeall#1{\xdef#1##1{#1{\noexpand\cite{##1}}}}
\def\cite#1{\each@rg\citer@nge{#1}}	% Variable No. of args, separated by

\def\each@rg#1#2{{\let\thecsname=#1\expandafter\first@rg#2,\end,}}
\def\first@rg#1,{\thecsname{#1}\apply@rg}	% each@ag is a general purpose
\def\apply@rg#1,{\ifx\end#1\let\next=\relax%	  variable no. of arg. macro.
\else,\thecsname{#1}\let\next=\apply@rg\fi\next}% args separated by commas

\def\citer@nge#1{\citedor@nge#1-\end-}	% Check for M-N range (M and N numbers)
\def\citer@ngeat#1\end-{#1}
\def\citedor@nge#1-#2-{\ifx\end#2\r@featspace#1 % Single argument
  \else\citel@@p{#1}{#2}\citer@ngeat\fi}	% M-N range of arguments
\def\citel@@p#1#2{\ifnum#1>#2{\errmessage{Reference range #1-#2\space is bad.}%
    \errhelp{If you cite a series of references by the notation M-N, then M and
    N must be integers, and N must be greater than or equal to M.}}\else%
 {\count0=#1\count1=#2\advance\count1 by1\relax\expandafter\r@fcite\the\count0,
  \loop\advance\count0 by1\relax%	  Loop from M to N
    \ifnum\count0<\count1,\expandafter\r@fcite\the\count0,%
  \repeat}\fi}

\def\r@featspace#1#2 {\r@fcite#1#2,}	% Eat spaces at beginning or end of arg
\def\r@fcite#1,{\ifuncit@d{#1}%		  Cite individual reference
    \newr@f{#1}%
    \expandafter\gdef\csname r@ftext\number\r@fcount\endcsname%
                     {\message{Reference #1 to be supplied.}%
                      \writer@f#1>>#1 to be supplied.\par}%
 \fi%
 \csname r@fnum#1\endcsname}
\def\ifuncit@d#1{\expandafter\ifx\csname r@fnum#1\endcsname\relax}%
\def\newr@f#1{\global\advance\r@fcount by1%
    \expandafter\xdef\csname r@fnum#1\endcsname{\number\r@fcount}}

\let\r@fis=\refis			% Save old \refis, redefine
\def\refis#1#2#3\par{\ifuncit@d{#1}%      Use two params #2 #3 to strip blank
   \newr@f{#1}%
   \w@rnwrite{Reference #1=\number\r@fcount\space is not cited up to now.}\fi%
  \expandafter\gdef\csname r@ftext\csname r@fnum#1\endcsname\endcsname%
  {\writer@f#1>>#2#3\par}}

\def\ignoreuncited{%   redefine \refis if ignoring uncited references
   \def\refis##1##2##3\par{\ifuncit@d{##1}%
    \else\expandafter\gdef\csname r@ftext\csname r@fnum##1\endcsname\endcsname%
     {\writer@f##1>>##2##3\par}\fi}}

\def\r@ferr{\endreferences\errmessage{I was expecting to see
\noexpand\endreferences before now;  I have inserted it here.}}
\let\r@ferences=\references
\def\references{\r@ferences\def\endmode{\r@ferr\par\endgroup}}

\let\endr@ferences=\endreferences
\def\endreferences{\r@fcurr=0%		  Save old \endreferences, redefine
  {\loop\ifnum\r@fcurr<\r@fcount%	  Loop over refnum and produce text
    \advance\r@fcurr by 1\relax\expandafter\r@fis\expandafter{\number\r@fcurr}%
    \csname r@ftext\number\r@fcurr\endcsname%
  \repeat}\gdef\r@ferr{}\endr@ferences}

% Save old \endpaper, redefine it to write parting message.

\let\r@fend=\endpaper\gdef\endpaper{\ifr@ffile
\immediate\write16{Cross References written on []\jobname.REF.}\fi\r@fend}

\catcode`@=12

\citeall\refto		% These macros will generate citations
\citeall\ref		%
\citeall\Ref		%

\def\pp{{\prime\prime}}

\def\a{{\alpha}}
\def\b{{\beta}}

\def\s{\sigma}
\def\half{{1 \over 2}}
\def\ra{{\rangle}}
\def\la{{\langle}}

\def\ih{{i \over \hbar}}

\def\E{{\cal E}}
\def\au{\underline \alpha}

\def\ria{{\rightarrow}}
\def\Tr{{\rm Tr}}

\def\H{{\cal H}}
\def\E{{\cal E}}

\def\la{\langle}
\def\ra{\rangle}
\def\ria{\rightarrow}

\def\s{{\sigma}}
\def\a{\alpha}
\def\b{\beta}
\def\e{\epsilon}
\def\U{\Upsilon}

\def\om{{\omega}}
\def\Tr{{\rm Tr}}
\def\ih{{ {i \over \hbar} }}

\def\au{{\underline \alpha}}
\def\bu{{\underline \beta}}
\def\pp{{\prime\prime}}
\def\id{{1 \!\! 1 }}

\centerline{\bf Somewhere in the Universe:}
\centerline{\bf Where is the Information Stored When Histories Decohere?}

\bigskip
\bigskip
\author J.J.Halliwell % \footnote{$^{\dag}$}{E-mail address: \jjh}
\vskip 0.2in
\affil
Theory Group
Blackett Laboratory
Imperial College
London SW7 2BZ
UK
\vskip 0.5in
\centerline {\rm Preprint Imperial/TP/98-99/29, quant-ph/9902008}
\vskip 0.1in
\centerline{\rm Second Revised version. July 26, 1999.}
\vskip 0.1in
\centerline {\rm Submitted to {\sl Physical Review D}}
\vskip 0.1in
\centerline {\rm  PACS numbers: 03.65.-w, 03.65.Bz, 98.80.Hw}
\vskip 0.3in

\abstract{ In the context of the decoherent histories approach to
quantum theory, we investigate the idea that decoherence is
connected with the storage of information about the decohering
system somewhere in the universe. The known connection between
decoherence and the existence of records is extended from the case
of pure initial states to mixed states, where it is shown that
records may still exist but are necessarily imperfect. We formulate 
an information-theoretic conjecture about decoherence due to an
environment: the number of bits required to describe a set of
decoherent histories is approximately equal to the number of bits of
information thrown away to the environment in the coarse-graining
process. This idea is verified in a simple model consisting of a
particle coupled to an environment that can store only one bit of
information. We explore the decoherence and information storage in
the quantum Brownian motion model, in which a particle trajectory is
decohered as a result of coupling to an environment of harmonic
oscillators in a thermal state. It is shown that the variables that
the environment naturally measures and stores information about are
non-local functions of time which are essentially the Fourier
components of the function $x(t)$ (describing the particle
trajectory). In particular, the records storing the information
about the Fourier modes are the positions and momenta of  the
environmental oscillators at the final time. We show that it is
possible to achieve decoherence even if there is only one oscillator
in the environment. The information count of the histories and
records in the environment add up according to our conjecture. These
results give quantitative content to the idea that decoherence is
related to ``information lost''. Some implications of these ideas
for quantum cosmology is discussed.
}

\endtopmatter  
\endpage

\head{\bf 1. Introduction}

The notion of decoherence plays an important role in discussions of
the foundations of quantum theory, particularly in investigations of
the emergence of classical behaviour [\cite{Har6,GeH2,JoZ,Zur2}].
Decoherence typically arises as a result of a coarse-graining scheme
-- dividing the system into subsystem and environment, for example,
and then tracing out the environment. Decoherence is then often
regarded as a kind of generalized measurement process: the
environment producing the decoherence ``measures'' the decohering
subsystem, and ``stores information'' about it. Indeed, it can be
argued that the physical significance of decoherence is that it
ensures the storage of information about the decohering system's
properties somewhere in the universe [\cite{GeH2,GeH3}]. 

These appealing ideas are frequently mentioned in the literature,
and some general theorems supporting them exist [\cite{GeH2,GeH3}].
However, it is probably fair to say that, despite the concrete
mathematical grip we now have on the notion of information, there is
still considerable scope for their development and implementation in
physically interesting models. This paper will focus on precisely
these issues, through two particular questions. First, when a system
decoheres as a result of coupling to an environment, how, in
practice, can the system's history be reconstructed by examining the
environment? That is, which properties of the environment carry the
information about the decohered system? Second, how is the {\it
amount} of information stored by the environment related to the
nature or degree of decoherence of the system?

We will address these issues in the context of the decoherent
histories approach to quantum theory [\cite{GeH1,GeH2,Gri,Omn,Har2,Hal1}].
(Other approaches to decoherence, such as Zurek's ``einselection''
approach [\cite{Zur1,Zur2}], related density matrix
approaches [\cite{JoZ}] or quantum state diffusion
[\cite{GiP,HaZ}], may be equally useful for analyzing these issues,
but will not be explored here.) In the decoherent histories
approach, probabilities are assigned to histories via the formula,
$$
p (\a_1, \a_2, \cdots ) =
{\rm Tr} \left( C_{\au} \rho C_{\au}^{\dag} \right)
\eqno(1.1)
$$
where $C_{\au} $ denotes a time-ordered string of projectors 
interspersed with unitary evolution,
$$
C_{\au}= 
P_{\a_n} e^{ -\ih H (t_n - t_{n-1}) }
P_{\a_{n-1}} e^{ -\ih H (t_2 - t_1) } \cdots P_{\a_1} 
\eqno(1.2)
$$
and $ \au $ denotes the string $\a_1, \a_2, \cdots \a_n $. Of
particular interest are sets of histories which satisfy the
condition of decoherence, which is that
decoherence functional
$$
D(\au,\au') = {\rm Tr} \left( C_{\au} \rho C_{\au'}^{\dag} \right)
\eqno(1.3)
$$
is zero when $\au \ne \au' $. Decoherence implies the weaker
condition that $ {\rm Re} D(\au,\au') = 0 $ for $\au \ne \au' $, and
this is equivalent to the requirement that the above
probabilities satisfy the probability sum rules.

But for us stronger condition of decoherence is the more interesting
one since it is related to the existence of records. In particular,
if the initial state is pure, there exist a
set of records at the final time $t_n $ which are perfectly
correlated with the alternatives $\a_1 \cdots \a_n $ at times $t_1
\cdots t_n $ [\cite{GeH2}]. This follows because, with a pure initial
state $| \Psi \ra$, the decoherence condition implies that
the states $ C_{\au} | \Psi \ra $ are an orthogonal set.
It is therefore possible to introduce a projection operator
$ R_{\bu} $ (which is generally not unique) such that
$$
R_{ \bu } C_{\au} | \Psi \ra = \delta_{\au \bu}
C_{\au} | \Psi \ra
\eqno(1.4)
$$
It follows that the extended histories characterized by the chain
$ R_{\bu} C_{\au} | \Psi \ra $ are decoherent, 
and one can assign a probability
to the histories $\au$
and the records $\bu$, given by
$$
p( \a_1, \a_2, \cdots \a_n ; \b_1, \b_2 \cdots \b_n )
= {\rm Tr} \left( R_{\b_1 \b_2 \cdots \b_n} C_{\au} \rho C_\au^{\dag} \right)
\eqno(1.5)
$$
This probability is then zero unless $\a_k = \b_k $ for all $k$, in
which case it is equal to the original probability  $ p(\a_1, \cdots
\a_n) $. Hence either the $\a$'s or the $\b$'s can be completely
summed out of Eq.(1.5) without changing the probability,
so the probability for the histories can be entirely
replaced by the probability for the records at a fixed moment of
time at the end of the history:
$$
p (\au ) = {\rm Tr} \left( R_{\au}  \rho (t_n ) \right)
= {\rm Tr} \left( C_{\au} \rho C_{\au}^{\dag} \right)
\eqno(1.6)
$$
Conversely, the existence of records $\b_1, \cdots \b_n $ at some
final time perfectly correlated with earlier alternatives $\a_1,
\cdots \a_n $ at $t_1, \cdots t_n $ implies
decoherence of the histories. This may be seen from
the relation
$$
D(\au, \au' ) = \sum_{\b_1 \cdots \b_n }
{\rm Tr} \left( R_{\b_1 \cdots \b_n } C_\au \rho C_{\au'}^{\dag} \right)
\eqno(1.7)
$$
Since each $\b_k $ is perfectly correlated with a unique
alternative $ \a_k $ at time $t_k $, the summand on the
right-hand side is zero unless $\a_k = \a_k'$ (although
note that, as we shall see later, a perfect correlation
of this type is generally possible only for a pure initial
state).

There is, therefore, a very general connection between decoherence
and the existence of records. From this point of view, the
decoherent histories approach is very much concerned with 
reconstructing possible past histories of the universe from records
at the present time, and then using these reconstructed pasts to
understand the correlations amongst the present records [\cite{Har7}].

The above results on the existence of records are very general, but
they do not give any idea as to how one can actually identify the
records in a given physical situation. How, for example, can one
identify the records  in the much-studied quantum Brownian motion
model, in which a large bath of oscillators in a thermal state
decohere a sequence of particle positions? In that model, the
environment in some sense ``measures'' the particle, so we expect the
records to be stored in the environment. Is it  in
practice possible to examine the environment at the final time and
explicitly reconstruct the past history of the particle?  Little
clue as to how one should do this is provided by the formal results
above. One main aim of this paper, as indicated
at the beginning of this Section, is therefore to show how to
actually find the records in the quantum Brownian motion model.

The second issue we will address, again as indicated, concerns the
{\it amount} of information stored in the records. Since the
environment is thought of as measuring and storing information about
the system, we expect there to be a quantitative connection between
the amount of information stored and the degree or nature of the
decoherence of the decohered system. What is the relevant measure of
the degree or ``amount'' of decoherence and how is it related to the
amount of information stored about the histories?

Thinking of decoherence via an environment as a generalized
measurement process, it is not difficult to see that the relevant
measure of the amount of decoherence is, loosely speaking, the
precision or width to within which the decoherent histories are
defined, or equivalently, the number of histories in the decoherent
set. (This issue, is, incidently, distinct from the question of the
degree of {\it approximate} decoherence, discussed below and
elsewhere [\cite{DoH,McE}]). To be more precise, a given set of
histories requires a certain number of bits of information to
describe it.  In the general account of histories and records given
above, suppose that the alternatives $\a_k$ run over $A$ values.
These could for example, be projections onto ranges of position that
partition the $x$-axis into $A$ different bins. Since $k=1, \cdots n$,
there are therefore $ A^n $ different histories, requiring
$\log_2 A^n $ bits of information to describe them. Clearly if these
histories are decoherent, the records they are correlated with must 
be able to store at least $ \log_2 A^n $ bits. For many practical
instances of decoherence, most of this information is stored in the
decohering environment, hence the environment Hilbert space must
have an information storage capacity large enough to accommodate
the information.

However, not all of this information needs to be stored in the
environment. This is because there can be a certain amount of
decoherence of histories even without coupling to an environment.
For example, the decoherence functional (1.3) is automatically
diagonal in the final alternatives $\a_n$ (because of the cyclic
property of the trace and the exclusive property of the projectors).
These alternatives don't require records since they exist at the
final time. More generally, for a system Hilbert space of dimension
$D_s$, since decoherence requires that the states $C_{\au} | \Psi
\ra $ must be orthogonal, there can in principle be a decoherent set
of as many as $D_s$ histories, without having to appeal to an
environment. (To reach this upper limit, however, requires that the
operators projected onto at each time are carefully chosen and
possibly not physically interesting). Hence, most generally, the
records consist of final projections onto both the distinguished
system and the environment. Furthermore, it is then clear that what
the environment stores information about is the {\it enhancement} in
the number of histories in a decoherent set when the system is
coupled to an environment. 

To be precise, return to the set of $A^n$ histories
descibed above. Since, as stated there is automatic decoherence 
of the $A$ final alternatives, it is the $A^{n-1}$ alternatives
at the $n-1$ earlier times that typically don't decohere without
an environment, and thus it is the records of these
$A^{n-1}$ alternatives that is stored in the environment.
If the labels of the records living in the
environment $\bu $ run over a total of $B$ values, we expect that a
necessary condition for decoherence is 
$$ 
B \ge A^{n-1}
\eqno(1.8)
$$  
This effectively mean that
there must be at least one register for each distinct history. 
If $ B < A^{n-1} $, each history cannot be uniquely correlated with a
record label $\b_k $, since there are not enough records, hence 
there will be no decoherence (in the pure state case).
Therefore, the amount of information stored in the environment
places an upper limit on the number of histories in the decoherent
set. Differently put, the environmental information storage capacity
limits the permissible amount of fine-graining of the system
histories consistent with decoherence.

The notion of the information of histories used here is clearly the
simplest one imaginable, but is actually sufficient for present
purposes. The general question of the assignment of information (or
entropy) to histories, and its relation to information storage in
the environment, is a very interesting one [\cite{IsL,Har3,HaB}],
but we will not go into it here. The possible difficulty is that a
Shannon-like information measure requires probabilities for
histories, but here we would like to discuss the logically prior
issue of decoherence, hence the existence of probabilities for
histories cannot be assumed. In any reasonable assignment of
information to histories, however, the value $\log_2 A^n $ will
typically arise as the maximum information, when the probabilities
for the histories are all equal, so here we are covering the worst
possible case.  This is actually appropriate to many of the
system--environment models studied in the literature, such as the
quantum Brownian motion model, where decoherence typically arises
for a fixed environment initial state with a wide class of system
initial states.  Decoherence is due in these models to the joint
system--environment dynamics and to the environmental initial state.
It does not depend very much on the system initial state, hence it
is appropriate to consider decoherence and information storage for a
variety of initial states.

Some comments on the nomenclature  ``information lost'' and
``records'' are in order. If the environment starts out in a pure
state, and its Hilbert space has dimension $D_e$, then its maximum
information storage capacity is $\log_2 D_e $ bits. Hence we would say
that the ``information lost'' to the environment is $\log_2 D_e $
bits, and we would also say that the records have $D_e$ different
possible states ({\it i.e.}, $B=D_e$, in the notation used above). If,
on the other hand, the environment is in a mixed state, the
``information lost'' to the environment can be greater than $\log_2
D_e $, since it also includes pre-exisiting uncertainty  (or
``information loss'') in the environment state. But the records
accessible by projections onto the environment still have $D_e$
different possible states, and in fact the number of {\it
distinguishable} environment states is often diminished in the
presence of a mixed state.  This will be discussed in more detail
later in the paper, but to be clear, the conjecture we will explore
is that in the case of both pure and mixed states, the amount of
decoherence is related to the ``information lost'' to the
environment, whether or not that information is accessoble through
projections onto the environment.

Note also that the above observation about the connection between
the information of histories and the size of the environment also
ignores the usual requirement of effective irreversibility  of
practical information storage. To store one bit in an effectively
irreversible fashion typically requires far more than one bit.
Hence we are not particularly concerned with practical information
storage (although that is ultimately in interesting issue to
pursue), rather the more fundamental question of the connection
between decoherence and maximum information storage.

The above arguments imply that in a
system-environment situation, if we throw away $N$ bits of
information by tracing out an environment of dimension $2^N $, we
could, in principle, find an enhancement in the number of histories
in a decoherent set by up to $2^N$. This means, for example, that if
we throw away just one bit, by coupling to a two-state system and
then tracing it out, we could increase the number of decoherent
histories by a factor of $2$. We will indeed produce such an
example. Crudely speaking, tracing out anything
ought to decohere something.

Another striking example is in the quantum Brownian motion model
[\cite{QBM}].
Conventional wisdom dictates that an environment of a large
number of oscillators is required to decohere histories of position
of a single point particle [\cite{JoZ}]. 
We will show, however, that even with an
environment of just {\it one} oscillator, decoherence of certain
variables describing the particle may be obtained. The variables in
question are defined non-locally in time, and are essentially the
Fourier modes of the particle's trajectory. This result then points
the way towards showing how the system's history may be recovered
from the oscillator states of a many-oscillator environment. This
simple example also sheds some light on the question of recurrences
and how it affects decoherence.

In addition to the issues of explicitly identifying the records, and
of finding a concrete connection between decoherence and information
storage, a third issue of relevance is the question of approximate
versus exact decoherence. In most realistic situations, decoherence
is only approximate. A reasonable conjecture is that an
approximately decoherent set of histories is in some sense close to
an exactly decoherent set, although it is generally difficult find
such exactly decoherent sets explicitly [\cite{App}]. Since decoherence is
related to the existence of records, one can imagine that the nature
of exact versus approximate decoherence could be better understood
by examining the nature of the records. To be more precise, since
records exist at a fixed moment of time at the end of the histories,
they are described by projections at just one time and they are
therefore trivially decoherent. It these records  are exactly
correlated with a set of alternatives in the past, those
alternatives would then be exactly decoherent. The extent to which
these alternatives are then ``close'' to a certain approximately
decoherent set of interest could then be assessed. Approximate
decoherence, may, for example, be approximate correlation of past
alternatives with an exactly decoherent set of records.  We
will have somewhat less to say about this issue than the other two,
but some comments can be made on the basis of the models examined,
and it will be taken up in more detail elsewhere.

In assessing the extent to which an environment ``measures''
or stores information about a system it is interacting with,
two different approaches suggest themselves.

The first, and simpler, approach is to examine explicit models of
the measurement process, in which the system of interest is coupled
to a measuring device specifically designed to become correlated
with the system in a particular way. In this way one can explicitly
see the information transfer from system to apparatus. However, one
can then also regard the apparatus as an environment for the system.
The apparatus states can then be traced out to produce decoherence
of certain system alternatives. One can then investigate the
connection between the decoherence of the system, and the extent to
which information about it is stored elsewhere.

The second approach is to do things the other way round. That is, to
start with a system coupled to an environment in a more general way,
which produces decoherence but less obviously corresponds to a
particular type of measurement. We can then ask whether, when
decoherence occurs, information about the system is in some sense
stored in the environment. In this paper, we will address these
issues in two models. 

We begin in Section 2 with a general discussion of records in the
case, not previously covered, in which the initial state is mixed.
It is argued that record-like projectors still exist, but their
correlation with past alternatives is necessarily imperfect. Records
in the case of decoherence by conservation are also discussed.

Section 3 concerns a model corresponding to the measurement process
which can also be used as an environment. It is a model for position
measurements which determine whether a particle has passed through a
series of spatial regions $R_1, R_2, \cdots $ at a series of times
$t_1, t_2, \cdots $. The measuring device consists of a series of
2-state systems localized to the regions $R_1, R_2, \cdots $, with
delta-functions in time, so the detectors are only on momentarily.
The coupling causes the 2-state system to flip from one state to the
other. Hence at the end of the history, one can discover whether the
particle was in $R_1$ at $t_1$, in $R_2 $ at $t_2 $ {\it etc.},  by
examining the state of the two-state systems. We thus obtain a very
simple model of the measurement process. We then trace out the
measuring devices and look for decoherence of the system alone.
Histories in which the position is specified to be in or not in
$R_1$, $R_2 $ etc, at times $t_1, t_2 $ are found to be exactly
decoherent. We thus find verification of our conjecture: the number
of bits required to describe a decoherent set of histories is equal
to the number of bits of information about the system stored in the
environment.

In Section 4 we consider the quantum Brownian motion model in
detail. It is first observed that classically, the response of  each
environmental oscillator in interacting with the particle trajectory
is to shift its final position and momentum by an amount
proportional to the Fourier modes of the trajectory. Essentially the
same story is shown to persist in the quantum case -- the shifted
position and momentum of the oscillators are the records storing
information about the Fourier modes. The information storage is
essentially perfect for a pure initial state for the environment,
but imperfect in the case of a mixed state. It is also seen that the
set of Fourier modes, in contrast to the particle trajectories,  are
in some sense the natural variables in which to discuss decoherence.
An elementary way of counting the number of histories in a
decoherent set is introduced, and this number is shown to
approximately coincide with the number of different possible record
states in the environment, in agreement with the conjecture.

Section 5 contains a discussion, including the implications of
some of these ideas for quantum cosmology.

This paper is builds very much on the connection between decoherence
and records in the decoherent histories approach, especially as put
forward by Gell-Mann and Hartle [\cite{GeH2,GeH3}],  although as
stated above, it is likely that other approaches to decoherence may
be amenable to a similar analysis. It is also partly inspired by
some of the ``It from Bit'' ideas initiated by Wheeler [\cite{Whe}]
and explored in detail by Caves [\cite{Cav}], Wooters [\cite{Woo}],
Zurek [\cite{Zur2,Zur4,Zur5}] and others [\cite{Zur3}]. One
particular motivation is the recent remark by Zurek [\cite{Zur3}],
that information-theoretic ideas have not been exploited to the
degree that they might. Indeed, before the advent of the decoherent
histories approach, it was Zurek who first spelled out the
connection between decoherence and information storage in the
environment [\cite{Zur4,Zur6}]. Some recent papers on the assigment
of information to histories by Hartle and Brun [\cite{HaB}],
Gell-Mann and Hartle [\cite{GeH3}],  and Isham and Linden 
[\cite{IsL}] have also been influential. Finally, is should be noted
that there has recently been a surge of interest in the subject of
{\it quantum} information but these interesting developments are not
very closely related to the present work, since we are interested in
the case in which the information stored by the environment is
essentially classical.

\head{\bf 2. Records in the Case of Mixed Initial States}

The connection between decoherence and records has been demonstrated
only in the case of a pure initial state [\cite{GeH2}].  Yet many
situations in which decoherence is studied, such as the quantum
Brownian motion  model, involve a thermal state for the environment
[\cite{QBM,DoH}], hence the overall initial state is mixed. In this
situation, the connection between decoherence and records needs to
be examined more carefully. There are two issues. First of all, to
determine whether records still exist in this case, that is, whether
it is still possible to add a record projector at the end of the
chain and preserve decoherence. Secondly, to work out the degree of
correlation between the records and the histories.

\subhead{\bf 2(A). Mixed Initial States}

We start from the observation that
a mixed state can always be regarded as the reduced density operator
of a pure state defined on an enlarged Hilbert space. Take,
for example, a mixed density operator of the form,
$$
\rho = \sum_n \ p_n \ | n \ra \la n |
\eqno(2.1)
$$
Suppose we enlarge the original Hilbert space $\H$
to $\H \otimes \tilde \H$, where $\tilde \H$ is an exact
copy of $\H $. Now on $\H \otimes \tilde \H$, we may
define the pure state,
$$
| \Psi \ra = \sum_n \ p_n^{\half} \ | n \ra \otimes \tilde {| n \ra}
\eqno(2.2)
$$
and it is readily seen that
$$
\rho = {\rm Tr}_{\tilde \H} \left( | \Psi \ra \la \Psi | \right)
\eqno(2.3)
$$
Of course, there are many different ways of regarding  finding a
pure state which reduces to a given mixed state in this way, but
this way is sufficient for illustrative purposes.

On the enlarged Hilbert space, the decoherence functional may be
written,
$$
D(\au, \au' ) = {\rm Tr} \left( C_{\au} \otimes \id \ | \Psi \ra \la \Psi|
\ C_{\au'}^{\dag} \otimes \id \right)
\eqno(2.4)
$$
where the trace is over $\H \otimes \tilde \H$. When there is
decoherence, the argument showing the existence of records may now
be repeated: there exist records at the final time perfectly
correlated with the alternatives $\au$.  The records exist, however,
on the enlarged Hilbert space. The probability of both the records
and the histories is
$$
p(\au, \bu ) = {\rm Tr} \left( \tilde R_{\bu}
\ C_{\au} \otimes \id \ | \Psi \ra \la \Psi|
\ C_{\au}^{\dag} \otimes \id \right)
\eqno(2.5)
$$
where $\tilde R_{\bu}$ is defined on  $\H \otimes \tilde \H$. 
It will generally not be possible to express this joint probability
in terms of states and projections on $\H$ alone. The projector on
the enlarged Hilbert space will generally be a sum of projectors of
the form $R \otimes Q $, where $R$ acts on $\H$ and $Q$ acts on
$\tilde \H$, and part of records will be contained in the projector
$Q$ on $\tilde H$.

Nevertheless, the existence of this joint probability distribution,
in which the addition of the records projector $\tilde R_{\bu }$ 
does not disturb the decoherence of the histories, permits us
to deduce the existence of an analagous formula on $\H$. For suppose
we coarse grain the record projector in such a  way that all
components $Q$ acting on $\tilde \H$ are replaced by the identity.
The decoherence of the histories is preserved since coarse graining
preserves decoherence. This implies that we may write down a joint
probability distribution of the form
$$
p(\au, \bu ) = {\rm Tr} \left( R_{\bu} C_{\au} \rho C_{\au}^{\dag}
\right)
\eqno(2.6)
$$
where everything is now defined on the original Hilbert space $\H $,
and $R_{\bu}$ is a projection operator.  Hence, given decoherence
in the case of a mixed initial state, we can always
add an extra projector $R_{\bu}$ at the end of the
chain without affecting decoherence. 

A less general, but perhaps more explicit discussion can be given by
appeal to the particular situations in which decoherence occurs. In
most physically interesting situations, the type of variables that
decohere, and that are correlated with records, is 
primarily determined by the
underlying Hamiltonian, and not by the initial state. The initial
state affects only the degree of decoherence and correlation. 
Suppose we first take as an initial state one of the pure states in
which the mixed initial state of interest is diagonal ({\it i.e.,}
one of the states $ | n \ra \la n | $ in the notation (2.1)).
Because the variables that decohere depend only on the Hamiltonian
we expect there to be {\it some} degree of decoherence for this state, and
record projectors may therefore be added at the end of the
histories, without affecting decoherence.  Now suppose we go from
this pure initial state to the mixed state. Since this is a coarse
graining, the extended histories including the records {\it continue
to be decoherent}. This implies that a formula of the type
(2.6) will exist and satisfy the probability sum rules.

The interpretation of the extra projector in (2.6) 
as an exact record makes sense
only on the enlarged Hilbert space. On coarse graining to the
original Hilbert space, the correlation between $\au$ and the
reduced set of records $\bu$ will generally be imperfect because we
have thrown away some of the information. This can be explicitly
shown as follows. Consider the conditional probability of the 
records $\bu$ given the past alternatives $\au$. This is given by
$$
\eqalignno{
p (\bu | \au ) &= { p (\au, \bu ) \over p (\au ) }
\cr
&= {\rm Tr} \left( R_{\bu} \ \rho_{eff} (\au) \right)
&(2.7) \cr }
$$
where
$$
\rho_{eff} (\au) = { C_{\au } \rho C_{\au}^{\dag }
\over {\rm Tr} \left( C_{\au} \rho C_{\au}^{\dag} \right) }
\eqno(2.8)
$$
A perfect correlation between the records and the past alternatives 
is assured only if $ p( \bu | \au ) = 1 $, which is possible
only if $\rho_{eff} ( \au ) $ is pure. If $\rho_{eff} ( \au ) $ is 
mixed, $ p (\bu | \au ) < 1 $, and the correlation is imperfect.

To see when $\rho_{eff} (\au ) $ is pure,
insert the diagonal form for $\rho$ in (2.8):
$$
C_{\au} \rho C_{\au}^{\dag} = \sum_n \ p_n \ C_{\au} | n \ra
\la n | C_{\au}^{\dag}
\eqno(2.9)
$$
$\rho_{eff} (\au) $ is therefore pure if and only if 
just one of
the terms in the sum on the right-hand side is non-zero.
This can only come about if
each $\au$ picks out a single value of $n$, 
so that for fixed $\au$,
$$
C_{\au} | n \ra = 0 
\eqno(2.10)
$$
except for just one value of $n$ corresponding to $\au$.
(The converse need not be
true, {\it i.e.}, the value of $n$ for which 
$C_{\au} | n \ra $ is non-zero may correspond to many values
of $\au$). The interesting case, however, is that in which
the fact that the initial state is mixed is essential for
decoherence, {\it i.e.}, 
there is no decoherence for the constituent pure 
initial states. In this case the states $C_{\au} | n \ra$ 
are generally not orthogonal,
$$
\la n | C_{\au'}^{\dag} C_{\au} | n \ra  \ne 0
\eqno(2.11)
$$
for $\au' \ne \au $.
This is incompatabile with (2.10). Hence, $\rho_{eff} (\au) $
can only be pure when there is decoherence for every
constituent pure component of the mixed initial state,
and in addition, the (rather special) condition (2.10)
is satisfied.

We therefore conclude the following: {\it when decoherence relies
on the impurity of the initial state, there are no records that are
perfectly correlated with the past alternatives}. When there is
decoherence for the constituent pure components of the initial
state, it will generally still be true that there are no perfect
records, except for the special types of histories for which
the condition (2.10) is satisfied.

Physically, the decay in quality of the records is no surprise. We
assign a mixed state when the system is subject to fluctuations
which are genuinely beyond our control so we average over them. For
example, all systems are subject to scattering by microwave
background radiation, and the scattered photons may subsequently
disappear beyond the horizon so are truly lost. This means that the
records themselves are in a mixed state, so suffer inescapable
fluctuations, and can therefore not be perfectly correlated with
anything.  However, although exact records are impossible, we might
reasonably expect to find final alternatives which are correlated
with past alternative to a good approximation. Indeed, we will find
this to be the case in the particular models we investigate.

The above arguments also illustrate why mixed
initial states tend to give better decoherence than pure ones ({\it
e.g.}, in the quantum Brownian motion model the decoherence improves
with increasing temperature of the thermal state of the
environment). By better decoherence, we mean that  more histories
decohere, or equivalently, that the  histories may be described more
finely without encountering interference effects. Earlier we put
forward the idea that the amount of decoherence is related to the
amount of information about the histories stored somewhere in the
universe. The more information stored the better the decoherence.
Since a mixed state may be regarded as the reduced density operator
of a pure state on an enlarged Hilbert space, it clearly represents,
compared to a pure state, an enhanced ability to store information,
since there is quite simply more Hilbert space available. Some of
that information is inaccessible from the original Hilbert space,
but that does not matter for the purposes of decoherence, which
depends only on the storage of information {\it somewhere}.

\subhead{\bf 2(B). Records in the Case of Decoherence by Conservation}

So far we have discussed decoherence arising from interaction with
an environment, and the associated information storage.  However,
decoherence of histories seemingly of a rather different nature comes
about when the alternatives characterizing the histories are exactly
conserved [\cite{HLM}]. This is an elementary property of the decoherence
functional -- the projectors commute with the unitary evolution
operators, so may all be moved up to the final time, where the
$P_{\a_k}$'s act on the $P_{\a_k'}$'s, and thus give diagonality of
the decoherence functional.  A more general notion which also gives
decoherence is {\it determinism} in the quantum theory. An example
is histories of projections onto large cells of phase space. These
projections have the property that under unitary evolution they
evolve (approximately) into another projection of identical type,
except that the centre of the phase space cell is shifted according
to the classical equations of motion [\cite{Omn}]. This approximate
determinism also guarantees (approximate) decoherence, for similar
reasons to the case of exact conservation. These mechanisms are
important in showing the emergent classicality of hydrodynamic
variables [\cite{Ana,BrH,BrHa1,Hal2}].

In these cases it is natural to again ask for the connection with
the existence of records, but the answer is almost trivial. Records
do not need to exist in a separate environment. The existence of
records is essentially the question of whether there exist
alternatives at the final moment of time which are perfectly
correlated with the alternatives describing the histories at earlier
times. Clearly the answer is yes in this case: histories of exactly
conserved quantities may always be expressed as projections at the
final moment of time, since the projectors may quite simply be moved
to the final time without changing anything. Each alternative at
each time is, in a sense, its own record. A similar story obtains in
the case of approximate determinism.

%\end

%\input c:/doc/macros/jnl.tex
%\input defs.tex

\head{\bf 3. A Two-State Environment}

In the Introduction, it was argued that decoherence is related to
information storage in the environment, and that the number of
histories in the decoherent set  is related to the amount of
information about the histories stored in the environment. Taken to
the extreme, this means that even an environment consisting of a
two-state system could potentially lead to decoherence of certain
system alternatives. In this Section, we will consider exactly such
an environment, and show that it provides an instructive model of
decoherence and information storage with exactly the expected
properties.

The system in question is taken to be a point particle coupled to a
two-state system environment via a coupling localized to a region of
space and which, for simplicity, acts only at a single moment of
time, $t=t_1$.  The two-state system has states $ | 0 \ra $, $ | 1
\ra $, with associated raising and lowering operators, $a$,
$a^{\dag}$, where
$$
a | 0 \ra = 0, \quad a | 1 \ra = | 0 \ra,
\quad a^{\dag} | 0 \ra = | 1 \ra, \quad a^{\dag} | 1 \ra = 0
\eqno(3.1)
$$
The Hamiltonian is
$$
H(t) = H_0 + \lambda \ \delta (t - t_1)
\ ( a + a^{\dag} ) \ \U (x) 
\eqno(3.2)
$$
where $H_0 = p^2 / 2m $.
Here, $\U (x)$ is a window function equal to $1$ in the interval $
[a,b]$ and zero outside it. Therefore, although we regard
the two-state system as an environment, it is also
a very simple model of
the measurement of position. If the two-state system is started out
in the state $ | 0 \ra $, it will flip to $ | 1 \ra $ if the
particle is in $[a,b]$ at time $t_1$, and remain in $ | 0 \ra $
otherwise. Hence by examining the state of the environment at any
time after $t_1$, we may recover one bit of information 
about the particle at time
$t_1$.

We assume that the initial state of the composite system is
$$
| \Psi_0 \ra = | \psi \ra \otimes | 0 \ra 
\eqno(3.3)
$$
It is convenient to introduce the eigenstates of $a + a^{\dag}$,
which are
$$
| \pm \ra = { 1 \over \sqrt{2} } \left( | 0 \ra \pm | 1 \ra \right)
\eqno(3.4)
$$
These we write as $ | s \ra $, where $ s = \pm 1 $, and we also have
$$
|0 \ra = {1 \over \sqrt{2}} \sum_s | s \ra, \quad
|1 \ra = {1 \over \sqrt{2}} \sum_s s | s \ra,
\eqno(3.5)
$$
The initial state may now be written
$$
| \Psi_0 \ra = { 1 \over \sqrt{2} } 
\sum_s \ | \psi \ra \otimes | s \ra 
\eqno(3.6)
$$

Now consider unitary evolution from $0$ to $t$, where $ 0 < t_1 < t
$. Since the product form (3.3) is preserved up to $t_1$, there is
no loss of generality in letting $t_1 \ria 0 $, and
$$
\eqalignno{
| \Psi \ra &= T \exp 
\left( - \ih \int_0^t dt' \ H (t') \right) | \Psi_0 \ra
\cr
&= {1 \over \sqrt{2} } \sum_s \ \exp \left( - \ih H t
\right) \ \exp \left( - \ih s \lambda \U (\hat x) \right)
\ | \psi \ra \otimes | s \ra
&(3.7) \cr}
$$
(where $T$ denotes time ordering).
The probability that the environment is then found in the state
$ | 1 \ra $ is given by
$$
\la \Psi | \ \left( \id_{\cal S} \otimes | 1 \ra \la 1 | \right) 
\ | \Psi \ra
= \int_a^b dx \ \sin^2 \left( { \lambda \over \hbar } \right)
\ | \psi (x) |^2
\eqno(3.8)
$$
This is the correct result of standard quantum measurement theory if
we choose the coupling $\lambda $ to be $ \lambda = \pi \hbar / 2 $,
so we now adopt this value.
With this value of $\lambda$, the second exponential in (3.7)
may be written,
$$
\exp \left( - {i \over 2 } \pi s \U ( \hat x) \right)
= ( 1 - \U (\hat x ) ) - is \U ( \hat x ) 
\eqno(3.9)
$$
since $\U $ is a window function, and therefore $\U^2 = \U $.
It follows that $ \U ( \hat x ) $ is also a projection operator
onto the region $[a,b]$, which we will denote by $P_y$, and
we will denote its negation $ 1 -\U (\hat x) $ by $ P_n$.

Now consider a history in which the system is hit by a projector
$P_{\a_1}$ at the initial time, and then a second projector
$P_{\a_2}$, at time $t$, both of these projectors acting only
on the particle, not the environment. Both will be projections
onto ranges of position, about which more below.
The decoherence functional
may be written,
$$
D(\a_1, \a_2 | \a_1', \a_2 ) = 
\la \Psi_{\a_1' \a_2} | \Psi_{\a_1 \a_2} \ra 
\eqno(3.10)
$$
where
$$
\eqalignno{
| \Psi_{\a_1 \a_2} \ra &=
P_{\a_2} \otimes \id_\E 
\ T \exp \left( - \ih \int_0^t dt' H(t') \right)
\ P_{\a_1} \otimes \id_\E
\ | \psi \ra \otimes | 0 \ra
\cr 
&=
{1 \over \sqrt{2} } \sum_s
\ P_{\a_2} \otimes \id_\E 
\ e^{ - \ih H t }
\ \left( P_n  - is P_y  \right)
\ P_{\a_1} | \psi \ra \otimes | s \ra
&(3.11) \cr}
$$
where the second line follows from (3.7) and (3.9). Summing over $s$
and using (3.5), we obtain,
$$
\eqalignno{
| \Psi_{\a_1 \a_2} \ra =
& \left( P_{\a_2} e^{ - \ih H t } P_n  P_{\a_1} | \psi \ra \right)
\otimes | 0 \ra
\cr
& \quad \quad \quad
-i \left( P_{\a_2} e^{ - \ih H t } P_y P_{\a_1} | \psi \ra \right)
\otimes | 1 \ra
&(3.12) \cr }
$$
In this expression the projectors $P_y$ and $P_n$ have come
entirely from the dynamics of the environment. It is therefore
reasonably clear that exact decoherence and a perfect
system--environment correlation is obtained if 
we choose the system projectors $P_{\a_1}$ to coincide
with $P_y$ and $P_n$. We have $P_n P_{\a_1} = 0$,
unless $\a_1 = n$, and $P_y P_{\a_1} = 0 $ unless $\a_1 =y $.
Therefore,
$$
| \Psi_{\a_1 \a_2} \ra =
\cases{
\left( P_{\a_2} e^{ - \ih H t } P_n | \psi \ra \right)
\otimes | 0 \ra, 
&if $ \a_1 = n$; \cr
-i \left( P_{\a_2} e^{ - \ih H t } P_y | \psi \ra \right)
\otimes | 1 \ra
&if $\a_1 = y$. 
\cr}
\eqno(3.13)
$$
from which the decoherence is easily seen. In these
expressions the projector $P_{\a_2}$ can be onto anything,
since decoherence in the final alternatives is always automatic.

An interesting alternative form of the decoherence functional
is its path integral form, derived directly from (3.7),
which is
$$
\eqalignno{
D(\au, \au' )  & =  \sum_{s}
\ \int_{\au} {\cal D } x \exp 
\left( \ih S [x(t) ] + {\pi i \over 2 }
s \U ( x(t_1) ) \right) \psi (x_0)
\cr
& \quad \times  \int_{\au'} {\cal D } y 
\exp \left( - \ih S [y(t) ] - {\pi i \over 2 }
s \U ( y(t_1) ) \right)
\psi^* (y_0 )
\cr
& =
\int_{\au} {\cal D} x \int_{\au'} {\cal D} y
\ \exp \left( \ih S [x(t)] - \ih S [y(t) ] \right)
\cr
& \quad \quad \quad
\times \ \cos \left( { \pi \over 2} 
[ \U (x(t_1) ) - \U (y(t_1) ) ] \right)
\ \psi (x_0 ) \psi^* (y_0 )
&(3.14) \cr}
$$
The cosine term plays the role of an influence functional, in that
it summarizes the effect of the environment. It may be seen that it
destroys interference between histories partitioned according to
whether they are in the region $[a,b]$ at time $t_1$, since it is
equal to $1$ if $x(t_1)$ and $y(t_1)$ are both either inside or
outside the region $[a,b]$, and is zero if one is inside and the
other outside.

Since the initial state of the whole system is pure, the existence
of exact decoherence means that there must exist records at the
final time. That is, we can add another projector $R_{\b}$ at the
final time and construct the probability $p(\a_1, \a_2, \b )$ where
$\b$ is perfectly correlated with $\a_1$. It is trivial to identify
the records -- they are clearly the states $| 0 \ra $, $|1 \ra $ of
the environment.  The record projectors $R_{\b}$ are
$$
R_{0} = \id_{\cal S} \otimes | 0 \ra \la 0 |,
\quad R_1 = \id_{\cal S} \otimes | 1 \ra \la 1 |
\eqno(3.15) 
$$
From (3.13) it is clear that
$$
R_{\b} | \Psi_{\a_1 \a_2} \ra = | \Psi_{\a_1 \a_2} \ra
$$
when $\a_1 = y $ and $\b=1$, or $\a_1 = n $ and $\b=0$,
with $R_{\b} | \Psi_{\a_1 \a_2} \ra = 0 $ otherwise.
There is therefore a perfect correlation between the records
and the past alternatives $\a_1$.
Essentially the same conclusions holds with different
choices of pure initial state. The main difference is
that the form of the record projectors change.

Turn now to the case in which the environment is in a mixed initial
state. First, we introduce a convenient notation in which (3.12) is
written
$$
| \Psi_{\a_1 \a_2} \ra = | \bar \psi_{\a_1 \a_2} \ra \otimes | 0 \ra
+ | \psi_{\a_1 \a_2} \ra \otimes | 1 \ra
\eqno(3.16)
$$
The joint probability of the histories and the records may
be written,
$$
p(\a_1, \a_2, \b) = \Tr \left( R_{\b}
| \Psi_{\a_1 \a_2} \ra  \la \Psi_{\a_1 \a_2} | \right)
\eqno(3.17)
$$
where 
$$
\eqalignno{
| \Psi_{\a_1 \a_2} \ra  \la \Psi_{\a_1 \a_2} | 
=
| \bar \psi_{\a_1 \a_2} \ra  \la \bar \psi_{\a_1 \a_2} | 
\otimes | 0 \ra \la 0 |
& +
| \psi_{\a_1 \a_2} \ra  \la  \psi_{\a_1 \a_2} | 
\otimes | 1 \ra \la 1 |
\cr
&+ \quad {\rm off-diagonal} \ {\rm terms}
&(3.18) \cr }
$$
The off-diagonal terms are irrevelant to both the discussion
of correlations and decoherence, since they make no contribution.
Eq.(3.18) is the case in which
the environment initial state is the pure state $| 0 \ra$,
and it shows very clearly the perfect correlation that exists
between the system histories and the environment states.
In particular, different system histories can be completely
distinguished by projecting onto the two orthogonal environment
states.
If the initial state instead were $ | 1 \ra $, then the result would
be similar to (3.18), but with the $ | 0 \ra $'s and $ | 1 \ra $'s
interchanged. It follows that if we take 
the environment initial state
to be the mixed state
$$
\rho_1 = a | 0 \ra \la 0 | + b | 1 \ra \la 1 |
\eqno(3.19)
$$
then the joint probability of the histories and the records
is
$$
p(\a_1, \a_2, \b)
= \Tr \left( R_{\b} \left(
| \bar \psi_{\a_1 \a_2} \ra  \la \bar \psi_{\a_1 \a_2} | 
\otimes \rho_1
+
| \psi_{\a_1 \a_2} \ra  \la \psi_{\a_1 \a_2} | 
\otimes \rho_2
\right) \right)
\eqno(3.20)
$$
where
$$
\rho_2 = b | 0 \ra \la 0 | + a | 1 \ra \la 1 |
\eqno(3.21)
$$

As described in Section 2, in the mixed state case the joint
probability (3.20) must necessarily indicate less than perfect
correlations between the records and the alternatives $\a_1$ in the
past. We can now see this in a different way. The point is that the
record projector needs to be able to unambiguously distinguish
between the different environment states the past alternatives are
perfectly correlated with.  This is possible in the pure state case,
where the alternatives $\a_1$ are perfectly correlated with the pair
of orthogonal pure states $ | 0 \ra$, $| 1 \ra$, and orthogonal pure
states are completely distinguishable. In the mixed state case, the
alternatives $\a_1$ become correlated with the two mixed states
$\rho_1$, $\rho_2$. These two states are {\it not} perfectly
distinguishable. There is no projection operator that can
unambiguously decide whether the environment is in state
$\rho_1$ or $\rho_2$.

The model therefore illustrates the generally expected features. We
can look at the environment and explicitly find the records. An
environment consisting of a two-state system leads to a decoherent
set of system histories enlarged by a factor of $2$ compared
to the set that decoherence without this environment.
Clearly if we attempted to consider more than two alternatives
$\a_1$, we would not expect decoherence. Decoherence is preserved as
we go to a mixed state (since there is decoherence for each
consituent pure state), and we see that the reason the records are
imperfectly correlated with past alternatives is due to the
impossibility of completely distinguishing between the mixed
environment states the system alternatives are correlated with.

This model can clearly be extended to more elaborate histories
involving an arbitrary number of alternatives at each moment of
time, and to an arbitrary number of times, but the essential ideas
have been established in this simple model. It is also perhaps of
interest to consider a slightly more realistic model of position
samplings involving a genuinely irreversible detector model that 
does not involve a delta-function in time. This has been considered
in Ref.[\cite{Hal3}].

%\end

%\input d:/macros/jnl.tex
%\input c:/doc/macros/jnl.tex
%\input defs.tex

\head{\bf 4. Decoherence and Information Storage in the
Quantum Brownian Motion Model}

In this Section we consider the question of how decoherence
is related to storage of information by the environment in
the quantum Brownian motion model. We begin with a brief
review of the model. Although standard material [\cite{QBM,CaL,FeV,
DoH,GeH2}],
it is presented at some length in parts since it will be necessary
to consider a modified version of the standard account later
on in this Section.

\subhead{\bf 4(A). The Quantum Brownian Motion Model}

We are concerned with the class of quantum Brownian models
consisting of a particle of large mass $M$ moving in a potential
$V(x)$ and linearly coupled to a  bath of harmonic oscillators.  The
total system is therefore described by the action,
$$
\eqalignno{
S_T [x(t), q_n(t)] = &\int dt \left[ \half M \dot x^2 - V(x) \right]
\cr & 
+ \sum_n \int dt \left[ \half m_n \dot q_n^2 - \half m_n \om_n^2 q_n^2
- c_n q_n x \right]
&(4.1) \cr}
$$
The decoherence functional has the form
$$
\eqalignno{
D ( \au, \au' ) =  \int_{\au}  {\cal D} x \int_{\au'} {\cal D} y
\int {\cal D} q_n {\cal D} r_n & \exp \left( \ih S_T [ x(t), q_n (t) ]
- \ih S_T [ y(t), r_n (t) ] \right) \ 
\cr
\times
& \rho_0 (x (0), y(0) ) \ \rho^{env}_0 ( q_n (0), r_n (0) )
&(4.2) \cr }
$$
where we have assumed a factored initial state. We will make the
standard assumption that the enviroment initial state is thermal
$$
\rho^{env}_0 (q_n, r_n ) = \prod_n 
\exp \left( - A (q_n^2 + r_n^2 ) + B q_n r_n \right)
\eqno(4.3) 
$$
where
$$
A = {m_n \om_n \over 2 \hbar} \coth \left( \hbar \om_n \b \right), \quad
B = { m_n \om_n \over \hbar \sinh \left( \hbar \om_n \b \right) } 
\eqno(4.4)
$$
and $ \b = 1 / k T $.
If the coarse
graining $\au$, $\au'$ 
does not involve the environment, it may be integrated
out, with the result
$$
D(\au, \au') = \int_{\au} {\cal D}x \int_{\au'} {\cal D}y \ 
\exp \left(
\ih S[x] - \ih S[y] \right) \ {\cal F} [ x(t), y(t) ] \ \rho ( x_0, y_0 )
\eqno(4.5)
$$
where
$$
S[x] =  \int dt \left[ \half M \dot x^2 - V(x) \right]
\eqno(4.6)
$$
and $ {\cal F} [ x(t), y(t) ] $ is the Feynman-Vernon influence
functional, 
$$
\eqalignno{
{\cal F} [ x(t), y(t) ] = \prod_n
& \int {\cal D} q_n {\cal D} r_n \ \rho^{env}_0 ( q_n (0), r_n (0) )
\cr
\times & \exp \left( \ih \int dt \left[ \half m_n \dot q_n^2 
- \half m_n \om_n^2 q_n^2 - c_n q_n x \right] \right)
\cr
\times & \exp \left( - \ih \int dt \left[ \half m_n \dot r_n^2 
- \half m_n \om_n^2 r_n^2 - c_n r_n y \right] \right)
&(4.7) \cr }
$$
The sum is over all paths for which meet, $q_n = r_n $, at the final
time and then there is an integral over $q_n$.

This expression may be evaluated by first using the standard
path integral for the propagator of a harmonic
oscillator in an external field,
$$
g( q^{\pp}_n, \tau | q'_n, 0 ) =
\int {\cal D} q_n \exp \left( \ih \int dt \left[ \half m_n \dot q_n^2 
- \half m_n \om_n^2 q_n^2 - c_n q_n x \right] \right)
\eqno(4.8)
$$
where the sum is over all paths $q_n(t)$ from $ q_n(0)=q_n'$ to
$ q_n (\tau ) = q^{\pp}_n $. The result is,
$$
g( q^{\pp}_n, \tau | q'_n, 0 ) =
\exp \left( \ih \left[ a {q^{\pp}_n}^2 + a {q'_n}^2 + b q^{\pp} q'
- c[x] q^{\pp} - d[x] q_n' - f[x] \right] \right)
\eqno(4.9)
$$
where
$$
\eqalignno{
a &= { m \om_n \cos \om_n \tau \over 2 \sin \om_n \tau }
&(4.10) \cr
b &= - { m \om_n \over \sin \om_n \tau}
&(4.11) \cr
c[x(t)] &= { c_n \over \sin \om_n \tau} \int_0^{\tau} dt
\ x(t) \ \sin \om_n t
&(4.12) \cr
d[x(t)] &= { c_n \over \sin \om_n \tau} \int_0^{\tau} dt
\ x(t) \ \sin \om_n (\tau - t)
&(4.13) \cr
f[x(t)] &= { c_n^2 \over m_n \om_n \sin \om_n \tau}
\int_0^{\tau} dt \int_0^t ds \ \ x(t) x(s) \
\sin \om_n (\tau - t) \ \sin \om_n s
&(4.14) \cr }
$$
Using these expressions, the initial state is folded in,
the final values of $q_n = r_n$ traced over, and the
influence functional obtained is then 
normally written in the form
$$
{\cal F} [x(t), y(t) ] = \exp \left( \ih W [x(t), y(t) ] \right)
\eqno(4.15)
$$
where, $W[x(t),y(t)]$ is influence functional phase, and has the form,
$$
\eqalignno{
W[x(t),y(t)] = & - 
\int_0^t ds \int_0^s ds' [ x(s) - y(s) ] \ \eta (s-s') \ [ x(s') + y(s') ]
\cr &
+ {i \over 2} \int_0^t ds \int_0^t ds' 
[ x(s) - y(s) ] \ \nu(s-s') \ [ x(s') - y(s') ] 
&(4.16) \cr }
$$
(In the imaginary part, the symmetry of $\nu (s-s') $ has been
used to write the two integrals over the same range, $[0,t]$,
and this will be exploited below).
The kernels $\eta(s)$ and $\nu(s)$ are defined by 
$$
\eta (s) = - \sum_n { c_n^2 \over 2 m_n \omega_n } \sin \omega_n s
\eqno(4.17)
$$
and
$$
\nu (s) = \sum_n { c_n^2 \over 2 m_n \omega_n } \coth ( \half \hbar
\omega_n \beta ) \cos \omega_n s
\eqno(4.18)
$$
They are commonly rewritten,
$$
\eqalignno{
\nu(s) = & \int_0^{\infty} { d \om \over \pi} I(\om) 
\coth\left( {\hbar \om \over 2 kT} \right) \cos \om s
&(4.19) \cr
\eta(s) = & { d \over ds} \gamma(s)
&(4.20) \cr }
$$
where
$$
\gamma(s) = 
\int_0^{\infty} { d \om \over \pi} { I(\om) \over \om} 
\cos \om s
\eqno(4.21)
$$
and $I(\om)$ is the spectral density
$$
I(\om) = \sum_n \delta (\om - \om_n) { \pi c_n^2 \over2 m_n \om_n}
\eqno(4.22)
$$

Typically, the spectral density is chosen to have the ohmic form
$$
I (\om) = M \gamma \om 
\ \exp \left(- { \om^2 \over \Lambda^2 } \right)
\eqno(4.23)
$$
Here, $\Lambda$ is a cut-off, which will generally be taken to be
very large.
We then find that
$$
\gamma(s) = M \gamma { \Lambda \over 2 \pi^{\half} } 
\ \exp \left( - {1 \over 4} \Lambda^2 s^2 \right)
\eqno(4.24)
$$
and thus when $\Lambda$ is very large, 
$$
\gamma(s) \approx M \gamma \delta(s)
\eqno(4.25)
$$
The noise kernel $ \eta (s) $ is non-local for large $\Lambda$, except in
the so-called Fokker-Planck limit, $ kT \ >> \ \hbar \Lambda$,  in which
case one has
$$
\nu(s) = { 2 M \gamma k T \over \hbar} \ \delta(s) 
\eqno(4.26)
$$

Decoherence of histories of positions typically arises
when there is essentially a continuum of oscillators
at high temperatures. For in this case,
$$
\left| {\cal F} [ x(t), y(t)] \right|
= \exp \left( - {2 M \gamma k T \over \hbar^2} \int dt (x-y)^2
\right)
\eqno(4.27)
$$
in the decoherence functional (4.5), hence the contribution from
widely differing paths $x(t)$, $y(t)$ is strongly supppressed. It
will be useful for what follows to spell out in more detail what
this means. Suppose that the coarse graining of the position
histories is chosen so that the histories are specified at each
moment of time up to a width $ \s $. This means that for pairs of
histories to be ``distinct'' in (4.27), $x$ and $y$ must differ by
at least $\s$. The decoherence condition, that (4.27) be very small,
is then a lower limit on the value of $\s$. If the time scale of the
entire history is $\tau$, the condition is
$
\s^2 >> { \hbar^2 / (2 M \gamma k T \tau) }
$
Hence, decoherence supplies a lower limit on the precision to within
which the histories of positions may be used in an essentially
classical way, without suffering interference effects. We can
discuss the number of decoherent histories in the set by confining
the particle's motion to a region of size $L$. Formally, this is of
course achieved by putting the system in a box, with the
accompanying complications. However, it is sufficient for our
purposes to restrict the particle's motion in a more approximate
way, by supposing that  the potential $V(x)$ becomes very large
outside the region, or by restricting to particle initial states
that have negligible support outside the region during the time
interval of interest. We can then say that for decoherence
to order $\s $ satisfying the above condition, 
the number of histories
in the decoherent set is of order $ L / \s  $.

Under more general conditions, the oscillatory and non-local nature
of the noise kernel $\eta (s)$ in $W$ makes decoherence of positions
at a series of times less obvious. This is not unrelated to the
presence of recurrences in the master equation. Take, for example,
the case of zero temperature and and a finite number of oscillators.
An arbitrary initial density operator might initially tend towards
diagonality in position, but over long periods of time, the
correlations ``lost'' to the environment will eventually come back,
and the density matrix will become off-diagonal.  In terms of the
decoherence functional, a set of decoherent histories defined in
terms of projections onto position at a sequence of times might lose
decoherence if the projections are spread out over a time-scale
comparable to the recurrence time. This is why it is necessary, at
least for decoherence of position, to take an essentially infinite
environment. We will see below, however, how this conclusion may be
modified.

\subhead{\bf 4(B). Decoherence of the Fourier Modes}

According to the general discussion in the Introduction, the
decoherence of histories of positions in the quantum Brownian motion
model means that there ought to exist records about the trajectories
$ x(t) $ somewhere in the  environment. We will now show how this
comes about. Important clues can be found from studying the classical
equations of motion of the environment of oscillators.
These are,
$$
m_n \ddot q_n + m_n \omega_n^2 q_n = - c_n x(t)
\eqno(4.28)
$$
The solution to this equation, with fixed $p_n (0)$, $q_n (0) $
is
$$
\eqalignno{
q_n (\tau ) &= q_n (0) \cos \omega_n \tau 
+ { p_n (0 ) \over m_n \omega_n } \sin \omega_n \tau 
\cr
& \quad \quad \quad - {c_n \over m_n \om_n } 
\int_0^\tau dt  \ x(t) \ \sin \omega_n (\tau -t) 
&(4.29) 
\cr
p_n (\tau ) &= { p_n (0 ) } \cos \omega_n \tau
- m_n \omega_n q_n (0) \sin \omega_n \tau 
\cr
& \quad \quad \quad - c_n  \int_0^\tau dt 
\ x(t) \ \cos \omega_n (\tau -t) 
&(4.30) \cr}
$$
where $p_n = m \dot q_n$.
From this solution, one can see that at the final time $\tau $,
the positions and momenta
of the environment of oscillators depend on the particle's trajectory $x(t)$
via the temporally non-local quantities
$$
\eqalignno{
X^s_n  &= \int_0^\tau dt  \ x(t) \ \sin \omega_n (\tau -t) 
&(4.31) \cr
X^c_n  &= \int_0^\tau dt  \ x(t) \ \cos \omega_n (\tau -t) 
&(4.32) \cr }
$$
Hence, classically, the final values of $p_n$ and $q_n$
are correlated with the variables 
$X^s_n$ and $X^c_n $ 
-- for given initial data for the environment, measurement of
the final data permits the determination of 
$X^s_n$ and $X^c_n $.

It now follows that, classically,  the {\it entire trajectory}
$x(t)$ for all $t$ may be recovered by using an infinite number of
oscillators, and by choosing the frequencies $\omega_n$
appropriately, since $X^s_n $ and $X^c_n $ are essentially the
Fourier components of the function $x(t)$ in its expansion on the
range $ [0,\tau]$. This is the key observation about how the
environment stores information about the system: each oscillator
measures a Fourier component of the trajectory.  We will demonstrate
that essentially the same story persists in the quantum theory. 

First, however, since we expect the non-local functions $X^s_n$,
$X^c_n$ to play a key role, let us explore their decoherence
properties. This is readily done in the decoherent histories
approach: the path integral form of the decoherence functional above
comfortably accommodates coarse grainings involving variables
defined non-locally in time. We calculate the decoherence functional
by summing over paths in which the functionals $X^s_n$, $X^c_n$ of
$x(t)$ are each constrained to lie in small widths, $\Delta_n $. This
can be achieved by inserting window functions, $\Upsilon_{\Delta}$,
which are $1$ inside a region of width $\Delta_n $ and zero outside.
Explicitly, the decoherence function has the form,
$$
\eqalignno{
D ( \au, \au' ) =  & \int {\cal D} x \int {\cal D} y
\exp \left( \ih S [ x(t) ] - \ih S [ y(t) ] + \ih W [x(t), y(t) ] \right)  
\ \rho_0 (x (0), y(0) ) 
\cr
\times
& \ \prod_n \ \Upsilon_{\Delta_n} ( X_n^s - \bar X_n^s )
\ \Upsilon_{\Delta_n} ( X_n^c - \bar X_n^c )
\ \Upsilon_{\Delta_n} ( Y_n^s - \bar Y_n^s )
\ \Upsilon_{\Delta_n} ( Y_n^c - \bar Y_n^c )
&(4.33) \cr }
$$
where $Y^s_n $ and $Y^c_n$ are
defined in terms of $y(s)$ exactly as in (4.31), (4.32),
and $\au $ now denotes the $\bar X_n^s$ and $\bar X_n^c $.
To see how well the variables $X^s_n$, $X^c_n$
decohere, we rewrite the influence functional in terms of them. 
Inserting the explicit form for $\nu (s)$, Eq. (4.18), and expanding
the factor $\cos \om_n (s-s') $, it is readily shown that 
$$ 
{\rm Im} W =   \sum_n { c_n^2 \over 4 m_n \om_n } 
\coth \left( { \hbar \omega_n \over 2 k T } \right) 
\left[ ( X^s_n - Y^s_n )^2 + (X^c_n - Y^c_n)^2 \right] 
\eqno(4.34) 
$$ 
Since the part of decoherence the functional governing decoherence
goes like $ \exp \left( - {\rm Im } W/ \hbar  \right) $ there is clearly
decoherence of the Fourier variables, provided that the widths
of their coarse graining are sufficiently large,
$$
\Delta^2_n 
{ c_n^2 \over m_n \hbar \om_n } 
\coth \left( { \hbar \omega_n \over 2 k T } \right)  \ >> \ 1
\eqno(4.35)
$$
Again this may be regarded
as a lower limit on the precision to within which the histories
may be defined.

An interesting feature of these expressions is that the oscillatory
functions of time are no longer present, since they have been
absorbed into the new non-local variables. It is therefore not
necessary to take an infinity of oscillators in the environment to
obtain decoherence, nor to take high temperatures. In particular,
there is a degree of decoherence, at any temperature, and {\it even
if there is only one oscillator in the environment}.

This last result is perhaps surprising, but it is in keeping with 
the idea put forward in the Introduction, which loosely speaking is
that tracing out {\it anything} coupled to the system ought to
produce decoherence of {\it something}.  The variables that decohere
are non-local in time,  and this is how they get round the old
problem of recurrences. Furthemore, the uncomplicated nature of the
decoherence of the Fourier modes, provides a useful alternative view
on decoherence of histories of positions in the case of low
temperatures, or finite environments, where the oscillatory and
non-local character of the noise kernels makes it difficult to get a
clear picture of the  decoherence of position histories. That is, we
regard the Fourier modes as in some sense more fundamental, and then
approximately reconstruct histories of positions from them. From now
on we will work entirely with particle trajectories characterized by
fixed values of the Fourier modes.

\subhead{\bf 4(C). System-Environment Correlations }

We turn now to the question of establishing the correlations between
the environment and system in the quantum case, and the consequent
decoherence. We have shown that, classically, the final values of
$q_n$ and $p_n$ are correlated with the Fourier components of the
particle's trajectory.  This can be established in the quantum
theory by considering a probability in which, in addition to
projecting onto the particle's trajectory at a series of times, we
also consider projections onto the final state of the environment.
In the quantum theory, one has to make a choice between projecting
onto final values of $q_n$ or $p_n$, or onto both using phase space
quasi-projectors. We first consider final states of the environment
characterized by fixed final values of $q_n$, denoted $q_n^{\pp}$. 
The general question is, given the probability $p(\au)$
for a decoherent set of histories, under what conditions  can one
introduce a record projector $R_{\bu}$ onto ranges of oscillator
positions at the end of the history, so that 
the probabilities for histories are
essentially undisturbed when the labels $\bu$ are suitably chosen?

We have shown that histories of the Fourier modes decohere
as long as they are coarse grained to a width $\Delta $, defined
above.
The probability for a set of histories plus records 
consisting of a projection $R_{\bu}$ onto ranges $\s$
of value of $q^{\pp}_n$ is,
$$
\eqalignno{
p(\au , \bu ) = \prod_n \int d q^{\pp}_n \ \Upsilon_\s ( q^{\pp}_n 
- {\bar q}_n )
\ & \int_{\au} {\cal D}x \int_{\au} {\cal
D} y 
\  \exp \left(
\ih S[x] - \ih S[y] \right) 
\cr
\ \times \ & {\cal F} [ x(t), y(t); \{ q_n^{\pp} \} ] \ \rho ( x_0, y_0 )
&(4.36) \cr}
$$
where $\{ q_n \} $ denotes the set of all oscillator coordinates.
$\Upsilon_{\s} $ is again a window function of width $\s$ which
implements the projection onto a range of $q^{\pp}_n$, centred
around ${\bar q}_n $ (which correspond to record labels $\bu$).. 
$\au $ denotes the paths of the particle in
configuration space specified by fixed values of 
the Fourier modes, as in (4.33).
The object ${\cal F} [ x(t), y(t); \{ q_n^{\pp}
\} ] $ is a generalized influence functional, given by the same path
integral expression, (4.7), but with the different  boundary
conditions that the final values of $q_n$ and $r_n$ are set to the
value $q_n^{\pp}$ (rather than summed over). Hence integrating 
${\cal F} [ x(t), y(t); \{ q_n^{\pp} \}] $  over all the
$q_n^{\pp}$'s, which is equivalent to letting $\s \ria \infty $ in
$\Upsilon_\s $, yields the usual influence functional, and hence
the original probability $p(\au)$. The question
is therefore to determine the smallest value of $\s$ for which
the probability $p(\au, \bu) $ is the same
as $p(\au) $, that is, the smallest value for which the integral
of $q^{\pp}_n$ over the range $\s$ is essentially equivalent
to integrating over an infinite range.

${\cal F} [ x(t), y(t); \{ q_n^{\pp} \}] $ may be written in terms 
of the propagator (4.9):
$$
\eqalignno{
{\cal F} [ x(t), y(t); \{ q_n^{\pp} \}] 
= \prod_n & \int dq_n' dr_n' \ \rho^{env}_0 (q_n, r_n ) 
\cr
& \times g( q^{\pp}_n, \tau | q'_n, 0 ) 
\ g^* ( q^{\pp}_n, \tau | r'_n, 0 ) 
&(4.37) \cr}
$$
The integrals are Gaussians, and at some length, one obtains
the result,
$$
\eqalignno{
{\cal F} & [ x(t), y(t); \{ q_n^{\pp} \}]  = 
\cr
\prod_n & \exp \left( - A \left( q^{\pp}_n - { d[x] \over b } \right)^2
- A \left( q^{\pp}_n - { d[y] \over b } \right)^2
+ B \left( q^{\pp}_n - { d[x] \over b } \right) 
\left( q^{\pp}_n - { d[y] \over b } \right)
\right)
\cr
\times & \exp \left( -\ih q^{\pp}_n \left( c[x] - c[y] 
+ \cos \om_n \tau  \left( d[x] - d [y] \right) \right) \right)
\cr
\times & \exp \left(
- { i \over 2 \hbar m_n \om_n } \sin \om_n \tau \cos \om_n \tau 
\left( d^2[x] - d^2[y] \right) - \ih \left( f[x] - f[y] \right)
\right)
&(4.38) \cr}
$$
where the coefficients $A$, $B$ are given by (4.4), and
$b$, $ c[x]$ and $d[x]$ are given by (4.10)--(4.14).
From these, and comparing with (4.31), we see that
$$
{ d[x] \over b } = 
- {c_n \over m_n \om_n } X_n^s \equiv - \tilde X_n^s
\eqno(4.39)
$$
Similarly, we also see that
$$
c[x] + \cos \om_n \tau  d[x] = c_n X_n^c
\eqno(4.40)
$$
Hence Eq.(4.38) may be rewritten
$$
\eqalignno{
{\cal F} & [ x(t), y(t); \{ q_n^{\pp} \}]  = 
\cr
\prod_n & \exp \left( - A \left( q^{\pp}_n  + \tilde X_n^s \right)^2
- A \left( q^{\pp}_n + \tilde  Y_n^s \right)^2
+ B \left( q^{\pp}_n + \tilde  X_n^s \right) 
\left( q^{\pp}_n + \tilde  Y_n^s \right)
\right)
\cr
\times & \exp \left( -\ih q^{\pp}_n c_n \left( 
X_n^c - Y_n^c \right) \right)
\cr
\times & \exp \left(
- { i \over 2 \hbar m_n \om_n } \sin \om_n \tau \cos \om_n \tau 
\left( d^2[x] - d^2[y] \right) - \ih \left( f[x] - f[y] \right)
\right)
&(4.41) \cr}
$$
As expected from the classical analysis, the first exponential in
Eq.(4.37) indicates that the oscillator coordinates $q^{\pp}_n$ 
are approximately correlated with the Fourier modes 
$ - \tilde X_n^s $.

To see more precisely the nature of the correlation,
note that the Gaussian in (4.41) may be rewritten, 
$$ 
\exp
\left( - (2A-B)  \left( q^{\pp}_n  + \half (\tilde X_n^s + \tilde
Y_n^s) \right)^2 - {1 \over 4} ( 2 A + B ) \left( \tilde X_n^s -
\tilde Y_n^s \right)^2 \right) 
\eqno(4.42)
$$ 
Clearly the second term in this exponential gives the
decoherence of the Fourier modes $\tilde X_n^s $ (since
this corresponds exactly to the usual imaginary part
of the influence functional phase (4.16) when the oscillator
coordinates are integrated out).
The decoherence width, $\tilde \Delta $ of these modes is
$$ 
\eqalignno{
\tilde \Delta & \equiv {c_n \over m_n \om_n } \Delta  
=( 2 A + B)^{-\half} 
\cr
& = \left( \tanh
\left( { \hbar \om_n \over 2 k T} \right) \right)^{-\half}
&(4.43) \cr}
$$
(in agreement with the earlier analysis, (4.35)). Hence a
projection onto a range of $q^{\pp}_n$ of {\it any} width
$\s$ can be added at the end of the histories without
affecting decoherence. In order to preserve the original
probabilities for the histories as much as possible,
however, the width $\s$ of the
record projection needs to satisfy 
$$  
\s > (2A-B)^{-\half} = \left(
\coth \left( { \hbar \om_n \over 2 k T} \right) \right)^{-\half} 
\eqno(4.44)
$$ 
for the integral to be equivalent to an integral over an infinite
range.

Generally the width $\s$ of the records $q^{\pp}_n$ will be greater
than the width $ \tilde \Delta $ of the decoherent histories of
Fourier modes, $\tilde X_n^s $. The correlation between them must
necessarily be imperfect, therefore, since the records alternatives,
being more coarsely defined, will not be able to completely
distinguish between different past history alternatives.
Differently, fixing a record alternative does not uniquely fix a
past history alternative, hence the conditional probability of the
histories given the records is less than one. Yet another way of
putting it is to say that, in a suitably chosen counting technique
(as we did after Eq.(4.27), for example), the number of records will
be {\it less} than the number of decoherent histories. The
imperfection of the records in the mixed state case can in fact be
understood already at a classical level. For even classically, the
amount of correlation between the phase space data of the
environment and the Fourier modes will be reduced if the 
environment is subject to thermal fluctuations.

In the case of a pure initial state for the environment, $B=0$, and
therefore $\s \sim \tilde \Delta $, and in this case we will have a
near-perfect correlation between the records and the histories (as
perfect as the degree of approximate decoherence, which is generally
extremely good).

General expectations are therefore confirmed: records exist
in the case of a pure initial state, with an almost
perfect correlation between the history alternatives and 
the records. In the mixed state case, records continue
to exist but with an imperfect correlation.

So far we have seen how projections onto ranges of the environmental
coordinates $q^{\pp}_n$ are correlated with the Fourier modes
$X_n^s$ describing the histories. This is, however, only a partial
description of the histories, since the variables $ X_n^c$, which
are in some sense complementary to $X_n^s$, 
also decohere. We expect these to
be correlated with the environmental momenta. 

To investigate
projections onto more general types of records, such as this,
at the final time, we
need to consider a more general type of influence functional in
which the paths summed over to obtain the influence functional (4.37)
are not constrained to meet at $q_n^{\pp}$, but may take
different values. This allows arbitrary states to be attached at the
final time. It is straightforward to show that this more general
influence functional is given by,
$$
\eqalignno{
{\cal F} & [ x(t), y(t); \{ q_n^{\pp}\} , \{ r_n^{\pp} \} ]  = 
\cr
\prod_n & \exp \left( - A \left( q^{\pp}_n  + \tilde X_n^s \right)^2
- A \left( r^{\pp}_n + \tilde  Y_n^s \right)^2
+ B \left( q^{\pp}_n + \tilde  X_n^s \right) 
\left( r^{\pp}_n + \tilde  Y_n^s \right)
\right)
\cr
\times & \exp \left( -\ih c_n \left( 
q^{\pp}_n X_n^c - r^{\pp}_n Y_n^c \right) \right)
\cr
\times & \exp \left(
- { i \over 2 \hbar m_n \om_n } \sin \om_n \tau \cos \om_n \tau 
\left( d^2[x] - d^2[y] \right) - \ih \left( f[x] - f[y] \right)
\right)
&(4.45) \cr}
$$
This object is, in fact, essentially just the thermal initial state, 
unitarily shifted in positions and momenta
by the classical equations of motion (4.29), (4.30) 
with vanishing
initial positions and momenta:
$$
{\cal F}  [ x(t), y(t); \{ q_n^{\pp}\} ,\{ r_n^{\pp} \} ]  = 
\la q_n^{\pp} | \ U (- \tilde X_n^s, - c_n X_n^c )
\ \rho^{env}_0 \ U^{\dag} (- \tilde Y_n^s, - c_n Y_n^c )
\ | r_n^{\pp} \ra
\eqno(4.46)
$$
(up to a possible phase).
Here, $U (q,p) $ represents the unitary displacement
operator in positions and momenta.
This result is not surprising for a
linear system.

Projections onto final momenta may be considered by Fourier
transforming with respect to both $ q_n^{\pp}$ and $r_n^{\pp}$.
In the zero temperature case, for which $B=0$, it is clear to see
what is going on. ${\cal F}$ has the form of the pure state density
operator for a coherent state of spatial width $ A^{-\half} $.
Fourier transform therefore leads to a state which has exactly the
same form; thus the discussion of decoherence and records is the
same as the previous case. The mixed state case will be similar.

Perhaps more useful and general is to combined the above two cases
and consider quasi-projectors onto the final values of the
environmental phase space data. Using (4.46), the explicit decoherence
functional for the situation involving any projection $R_{\bu}$
onto environment states at the final time is
$$
\eqalignno{
D(\au ,\au', \bu ) = &
\int_{\au} {\cal D}x \int_{\au'} {\cal D} y 
\  \exp \left(
\ih S[x] - \ih S[y] \right) \ \rho_0 (x_0, y_0 )
\cr
\ & \times \ 
{\rm Tr} \left( R_{\bu} \ \ U (- \tilde X_n^s, - c_n X_n^c )
\ \rho^{env}_0 \ U^{\dag} (- \tilde Y_n^s, - c_n Y_n^c )
\right)
&(4.47) \cr}
$$
where the trace is over the environment Hilbert space. 
It is clear that decoherence and the probabilities for histories
are not disturbed if the records projectors $R_{\bu}$ are taken
to be phase space quasi-projectors onto suitable large regions
of phase space, and the discussion is again very similar, so
need not be spelled out in detail.

To summarize, the classical analysis shows that the Fourier modes of
the particle trajectories are correlated with the final values of
the phase space data of the environment at the final time. We have
shown that essentially the same story persists in the quantum
theory. For the zero temperature case, the record projectors need to
be wide enough to beat quantum fluctuations. For finite temperature,
they need in addition to beat the thermal fluctuations, and the
correlation between the records and the history alternatives is then
less than perfect, in accordance with general expectations.

It is also worth noting that the discovered correlation of the
final phase space data with the Fourier modes of the environment
means that the environment effectively performs a so-called
spectral measurement of the particle's trajectory. Measurements
of this type have previously been investigated by Mensky
in the context of the path integral approach to continuous
quantum measurement [\cite{Men}].

\subhead{\bf 4(D). Information Count}

We may now check that, as asserted at the beginning of the paper,
the amount of decoherence is related to the amount of information
thrown away. That is, the number of histories in the decoherent set
is approximately the same as the number of states thrown away to the
environment. We will consider the most general case considered
above, in which the system histories are characterized by $ X_n^s$,
$X_n^c$, and the records are phase space projectors onto the
environmental oscillators.

Consider first the case of zero temperature. Since the variables we
are dealing with are continuous and the Hilbert spaces infinite
dimensional, we need to make some artificial restrictions in order
to do any counting. Hence, as earlier in this section, lets us
restrict the dynamics of the distinguished particle to a spatial
region of size $L$. The Fourier variables (4.31), (4.32), are
therefore restricted to a region of size of order $L \tau$.

For decoherence, the widths $\Delta_n$ of the Fourier variables must
satisfy the inequality (4.35), which for $T=0$ reads
$ \Delta^2_n > m_n \hbar \omega_n / c_n^2 $. The histories of
the two types of
Fourier variables, $X_n^s $ and $X_n^c$ are therefore each defined up
to order $\Delta_n$, satisfying this restriction, and
there are of order $ L \tau / \Delta_n $ decoherent histories
of the variables $X_n^s$ and the same number of the variables
 $X_n^c$. Hence,
for each mode $n$,
the total number of histories $N_d$ in the decoherent set is 
$$
N_d = \left( { L \tau \over \Delta_n } \right)^2
\ < \
{ c_n^2 L^2 \tau^2 \over m \hbar \omega_n }
\eqno(4.48)
$$

Now consider the environment states for each mode $n$.  Each
environment mode starts out centred around $q_n = 0 = p_n$, and as a
result of interacting with the system, is displaced in $q_n$ and
$p_n$ by the amounts (4.29), (4.30).  (A partially classical
analysis suffices since the system is linear). Since $x(t)$ is
assumed to be restricted to a region of size $L$, $q_n$ will range
over a region with size of order $ c_n L \tau / m_n \om_n $, and
$p_n$ will range over a region of size $ c_n L \tau $. $q_n$ and
$p_n$ therefore  range over a phase space volume of size $ c_n^2 L^2
\tau^2 / m_n \om_n $.  The number of distinct
environment states, for each mode $n$,
corresponding to this
phase space volume is therefore given by
$$
N_{\e} = { c_n^2 L^2 \tau^2 \over m_n \hbar \om_n}
\eqno(4.49)
$$
which coincides with (4.48). This is therefore the desired result:
the number of distinct states of the environment thrown away in the
coarse graining process is equal to the upper limit on the number of
histories in the decoherent set of histories. Differently put, the
record of each individual history of the Fourier variables is stored
in a single phase space cell of an environment oscillator.

In the case of a thermal environment with $T \ne 0 $, decoherence is
improved so that, from (4.35), 
the number of histories in the decoherent set has a
larger upper limit:
$$
N_d = \left( { L \tau \over \Delta_n } \right)^2
\ < \
{ c_n^2 L^2 \tau^2 \over m \hbar \omega_n } \coth \left( { \hbar
\om_n \over 2 k T } \right)
\eqno(4.50)
$$
The effect of thermal fluctuations on the environment states is,
from one point of view, to reduce the number of distinguishable
states, since the elementary phase space cells are effectively
increased in size from $\hbar $ to $\hbar \coth ( { \hbar \om_n / 2
k T } ) $ in a thermal state. That is, the number of distinct  {\it
accessible} records in the environment is reduced. However, as
discussed in Section 2, a mixed environment state can be regarded as
a pure state on an enlarged environment Hilbert space, much of which
is inaccessible, and it is from the perspective of this enlarged
environment Hilbert space that we expect to understand the 
connection between decoherence and information loss.

There are then a number of ways of understanding how much
information is stored in the enlarged Hilbert space. For example, we
can regard the smearing of the environment phase space cells from
$\hbar $ to $\hbar \coth ( { \hbar \om_m / 2 k T } ) $ as meaning
that the environment is actually in one of a number $\coth ( { \hbar
\om_n / 2 k T } ) $  of $\hbar$-sized phase space cells, but the
information as to exactly which of those cells it occupies
is stored in the
inaccessible part of the enlarged Hilbert space. This indicates that
the mixed state case, regarded as a pure state on an enlarged
Hilbert space, has its information storage capacity enhanced by a
factor of $\coth ( { \hbar \om_n / 2 k T } ) $ compared to the $T=0$
case. Hence Eq.(4.49), the information storage capacity of  one mode
of the environment in the $T=0$ case, is multiplied by the factor, 
$\coth ( { \hbar \om_n / 2 k T } ) $, and we obtain agreement with
Eq.(4.50). That is, in the mixed initial state case also, the number
of histories in the decoherent set is approximately the same as the
maximum number of states storing information about the histories.

Another way of understanding the increased information storage in
the mixed state case is to consider the von Neumann entropy $ S = -
{\rm Tr} \left( \rho \ln \rho \right) $ of the environment. Loosely
speaking, in going from a pure to a mixed state, the number of
states available for information storage is increased by $e^S $. It
is well-known that the entropy of a harmonic oscillator in a thermal
state is of order $\ln ( k T / \hbar \omega ) $, hence the
information storage enhancement factor is of order $ k T / \hbar
\omega_n $, for large $T$. This agrees with the $\coth ( \hbar \om_n
/ 2 k T ) $ factor deduced above in the limit of high temperatures.
It does not generally agree at lower temperatures, although this is
not surprising  since measures of uncertainty or information loss in
quantum theory are dependent on the particular dynamical variables
of interest. (Since a thermal state is diagonal in energy, the von
Neumann entropy may be regarded as a measure of uncertainty in
energy, which  will generally not be the same as the phase space
uncertainty used above). Nevertheless, these two arguments are
sufficient for it to be seen that the degree of decoherence (4.50)
may be related to information lost to the environment in the mixed
state case.

\subhead{\bf 4(E). Exact and Approximate Decoherence}

Finally, we may make some elementary remarks about approximate
decoherence. Inserting (4.46) in the expression for the joint
probability of the histories and the records, we obtain
the particularly simple expression,
$$
\eqalignno{
p(\au , \bu ) = &
\int_{\au} {\cal D}x \int_{\au} {\cal D} y 
\  \exp \left(
\ih S[x] - \ih S[y] \right) \ \rho_0 (x_0, y_0 )
\cr
\ & \times \ 
{\rm Tr} \left( R_{\bu} \ \ U (- \tilde X_n^s, - c_n X_n^c )
\ \rho^{env}_0 \ U^{\dag} (- \tilde Y_n^s, - c_n Y_n^c )
\right)
&(4.51) \cr}
$$
where the trace is over the environment Hilbert space. For
simplicitly take the environment initial state to be pure, so it is
the ground state of the harmonic oscillator $| 0 \ra $.  The unitary
displacement operators then turn it into standard coherent states. 

The issue of exact decoherence or exact correlation
of the records with the histories, is then the question of
finding a coarse graining of the Fourier modes,
$ X_n^s $ and $X_n^c $ which effectively brings the
coherent states
$$
\ \ U (- \tilde X_n^s, - c_n X_n^c ) | 0 \ra
$$
into an orthogonal set of states. It is well-known that given the
coherent states $ |p,q, \ra $, which are overcomplete, a complete
set of states may be found by restricting $p,q$ to discrete values
lying on a regular lattice, and this is clearly implementable by
suitable coarse graining of the Fourier modes [\cite{Bac}]. The resulting
states, however, are not orthogonal. The orthogonalization process
may not be straightforward to carry out. More signficantly,
it is by no means clear that a coarse graining of the Fourier modes
is possible which puts this orthogonalization process into
effect. The issue of finding an exactly decoherent set of
histories which is close to the approximately decoherent sets
discussed in this section therefore remains open.

%\end

\head{\bf 5. Summary and Discussion}

We have obtained a number of results concerning the connection
between decoherence, information loss and the existence of records.
The existing basic result that we have very much built on is the
fact that decoherence with a pure initial state implies the
existence of records [\cite{GeH2}], {\it i.e.}, alternatives that
may be added to the end of the histories that are perfectly
correlated with the past alternatives. Our main aim was to explore
the connection between decoherence and {\it physical} information
storage in the case of decoherence due to an environment. The main
results may broadly be summarized as follows.

\noindent{\bf 1.} The discussion of records in the decoherent histories
approach has been extended to the case of mixed initial states, both
in the general results of Section 2, and the explicit models of
Sections 3 and 4.

\noindent{\bf 2.} In the quantum Brownian motion model the records
carrying information about the distinguished particle's trajectory
have been explicitly identified.

\noindent{\bf 3.} We have formulated a concrete conjecture
concerning the amount of  decoherence and the information lost to
the environment, and proved it in some important specific cases.
This gives substance to the old idea that decoherence is related to
information loss.

This last result also indicates how decoherence
conditions in practical models can be interpreted.
The commonly used decoherence condition,
$$
\exp \left( - { {\rm Im} W [x(t), y(t) ] \over \hbar } \right) << 1
\eqno(5.1)
$$ 
(where $W$ is the
influence functional), is normally physically interpreted as meaning
that interference between trajectories $ x(t) $, $y(t)$ is very
small, and therefore that probabilities may be assigned to these
histories. That is, the condition (5.1) puts a lower
limit on the degree to which the histories may be fine-grained 
without interference effects becoming significant.

To assign probabilities, one requires only the condition of
consistency, $ {\rm Re} D (\au, \au') = 0 $ for $\au \ne \au'$, 
whereas the condition (5.1) corresponds to the stronger condition of
(approximate) decoherence,  $ D (\au, \au') \approx  0$ for $\au \ne
\au'$, which surely merits a stronger interpretation.  The physical
meaning of decoherence  is that it implies the existence of records,
as discussed in Sections 1 and 2 (and in Ref.[\cite{GeH2}]). 
Consistency alone does not guarantee this. The decoherence condition
(5.1) should therefore be thought of in terms of the records, rather
than just in terms of interference and the assignment of
probabilities. In this paper we have effectively shown that {\it the
decoherence condition is a reflection of the  information storage
capacity of the environment.} That is, it is a lower limit on the
degree to which the histories may be fine-grained without the
information storage capacity of the environment being exceeded.

Some of the issues considered in this paper shed some light on an
old problem with decoherence in the context of quantum cosmology, 
which is how to choose the division of the universe into ``system''
and ``environment''.  On perusing the literature on decoherence via
tracing out an environment, one can find papers in which a matter
field is traced out to produce decoherence of the gravitational
field in quantum cosmology [\cite{Hal4}]. On the other hand, one can find
other papers in which the gravitational field is regarded as a
decohering environment for matter, since it is clearly the universal
environment [\cite{Ana2}]. 
Which is correct? One man's system is another man's
environment, at least,  from the point of view of published papers
on the subject, 

The case of decohering the gravitational field is an interesting
one, since the gravitational field is undeniably classical in  all
physical situations that can be checked observationally, so there is
a strong incentive to discover the mechanism by which it becomes
classical from the assumed underlying quantum gravity theory. On the
other hand, in a certain sense we {\it never} actually measure the
gravitational field itself. What we actually measure are the changes
of motion of matter that we interpret as being due to an underlying
gravitational field. From that point of view, nothing is really lost
by tracing out the gravitational field since it is never really
actually observed.

The ideas discussed in this paper  perhaps offer some resolution to
the dilemma over the choice of ``system'' and ``environment''. As we
have seen in a number of situations, decoherence is intimately
connected with the existence of records at the final moment of time
that are correlated with alternatives in the past. Furthermore, as
we saw in the analysis of the quantum Brownian motion model, the
records can be stored in the decohering environment, and by
inspecting them at the final time we can recover the past  history
of the system. What the decoherence of the quantum Brownian particle
by a thermal environment means, therefore, is that the history of
the Brownian particle may be recovered by examining the thermal
environment. Similarly, the decoherence of a gravitational field by
a decohering matter field environment means that we can recover the
history of the gravitational field by examining the matter field at
late times, which is indeed exactly what is done in cosmology. From
a practical point of view therefore, the significance of decoherence
is that it ensures a correlation between present records and past
events.

\head{\bf Acknowledgements}

I am very grateful to Todd Brun, Jim Hartle, Jason Twamley  and
Wojtek Zurek for useful conversations. Todd Brun is particularly
thanked for pointing out a number of errors in the original
manuscript.

\references

\def\pr{{\sl Phys.Rev.}}
\def\jmp{{\sl J.Math.Phys.}}

\refis{Ana} C.Anastopoulos,
``Hydrodynamic equations from quantum field theory",
preprint gr-qc/9805074 (1998).

\refis{Ana2} C.Anastopoulos, {\sl Phys.Rev.} {\bf D54}, 1600 (1996);
B.Kay, {\sl Class.Quant.Grav.} {\bf 15}, L89 (1998).

\refis{App} See, for example, H.F.Dowker and A.Kent, {\sl
J.Stat.Phys.} {\bf 82}, 1575 (1996). These authors actually made the
weaker conjecture that an approximately {\it consistent} set of
histories  ({\it i.e.}, ${\rm Re} D(\a, \a') \approx 0 $ for $\a \ne
\a' $) are close to an exactly consistent set.
R.Omn\`es (private communication) has pointed out that this is
easily proved in some models: the real part of the decoherence
functional is typically highly oscillatory as a function of the
times of the projections, hence if it is close to zero, it can be
made exactly zero by  a small change in the times. The stronger
conjecture  involving decoherence (rather than just consistency) is
very plausible but yet unproven in general.

\refis{Bac} See, for example, H.Bacry, 
A.Grossman and J.Zak, \pr {\bf B12}, 1118 (1975).
% Proof of completeness of lattice states in the kq representation.

\refis{BrH} T.Brun and J.J.Halliwell, 
{\sl Phys.Rev.} {\bf 54}, 2899 (1996).
%``Decoherence of Hydrodynamic Histories: A Simple Spin Model''

\refis{BrHa1} T.Brun and J.B.Hartle, ``Hydrodynamic Equations from
Decoherent Histories'', Santa Barbara preprint (in preparation).

\refis{CaL} A.Caldeira and A.Leggett, {\sl Physica} {\bf 121A}, 587 (1983).

\refis{Cav} C.Caves and C.A.Fuchs, quant-ph/9601025;
% Quantum information: How much information in a state vector?
C.Caves, in {\it Physical Origins of Time Asymmetry},
edited by  J.J.Halliwell, J.Perez-Mercader and W.Zurek (Cambridge
University Press, Cambridge, 1994). 

\refis{DoH} H. F. Dowker and J. J. Halliwell, {\sl Phys. Rev.} {\bf
D46}, 1580 (1992).
%{\it Quantum Mechanics of History: The Decoherence Functional in
%Quantum Mechanics.}

\refis{FeV} R.P.Feynman and F.L.Vernon, 
{\sl Ann. Phys.} {\bf 24}, 118 (1963).

\refis{GeH1} M.Gell-Mann and J.B.Hartle, in {\it Complexity, Entropy 
and the Physics of Information, SFI Studies in the Sciences of Complexity},
Vol. VIII, W. Zurek (ed.) (Addison Wesley, Reading, 1990); and in
{\it Proceedings of the Third International Symposium on the Foundations of
Quantum Mechanics in the Light of New Technology}, S. Kobayashi, H. Ezawa,
Y. Murayama and S. Nomura (eds.) (Physical Society of Japan, Tokyo, 1990).
%{\it Quantum Mechanics in the Light of Quantum Cosmology.}

\refis{GeH2} M.Gell-Mann and J.B.Hartle, {\sl Phys.Rev.} {\bf D47},
3345 (1993).
%{\it Classical Equations for Quantum Systems}.

\refis{GeH3} M.Gell-Mann and J.B.Hartle, 
in {\it Quantum Classical Correspondence: Proceedings of the
4th Drexel Symposium on Quantum Nonintegrability},
edited by D.H.Feng and B.L.Hu (International Press, 1997).
% gr-qc/9509054 (1995). strong deocherence.

%\refis{GiR} I.Giardina and A.Rimini, {\sl Found.Phys.} {\bf 26}, 
%973 (1996).
% ON THE EXISTENCE OF INEQUIVALENT QUASIDETERMINISTIC DOMAINS

\refis{GiP} N.Gisin and I.C. Percival, {\sl J.Phys.} {\bf A25},
5677 (1992); {\bf A26}, 2233 (1993); {\bf A26}, 2245 (1993);
{\sl Phys. Lett.} {\bf A167}, 315 (1992).

\refis{Gri} R.B.Griffiths, {\sl J.Stat.Phys.} {\bf 36}, 219 (1984);
{\sl Phys.Rev.Lett.} {\bf 70}, 2201 (1993).

%\refis{Gri2} R.B.Griffiths, {\sl Phys.Rev.} {\bf A54}, 2759 (1996);
%{\bf A57}, 1604 (1998).

\refis{Hal1} J.J.Halliwell, in {\it Fundamental Problems in Quantum
Theory},  edited by D.Greenberger and A.Zeilinger, Annals of the New
York Academy of Sciences, Vol 775, 726 (1994).

\refis{Hal2} J.J.Halliwell, {\sl Phys.Rev.} {\bf D58}, 105015 (1998);
quant-ph/9905094.
%``Decoherent Histories and Hydrodynamic Equations'',
% Preprint IC/TP/97-98/50. 

\refis{Hal3} J.J.Halliwell, ``Arrival Times in Quantum Theory from an Irreversible
Detector Model'', quant-ph/9805057, Imperial College preprint
TP/97-98/49.

\refis{Hal4} J.J.Halliwell, \pr {\bf D39}, 2912, 1989;
%``Decoherence in Quantum Cosmology", 
T.Padmanabhan, \pr, {\bf D39}, 2924 (1989);
%Decoherence in the density matrix describing quantum three-geometries and the
%emergence of classical spacetime.
C.Kiefer, {\sl Class.Quant.Grav.} {\bf 6}, 561 (1989).
%Continuous measurement of intrinsic time by fermions.

\refis{HaZ} J.J.Halliwell and A.Zoupas, {\sl Phys.Rev.}
{\bf D52}, 7294 (1995); {\bf D55}, 4697 (1997).

%\refis{Har1} J.B.Hartle, in {\it Quantum Cosmology and Baby
%Universes}, S.Coleman, J.Hartle, T.Piran and S.Weinberg (eds.)
%(World Scientific, Singapore, 1991).  {\it The Quantum Mechanics of
%Cosmology.}

\refis{Har3} J.B.Hartle, {\sl Phys.Rev.} {\bf D51}, 1800 (1995).
%Spacetime Information

\refis{Har2} J.B.Hartle, in {\it Proceedings of the 1992 Les Houches Summer
School, Gravitation et Quantifications}, edited by B.Julia and J.Zinn-Justin 
(Elsevier Science B.V., 1995).
%{\it Spacetime Quantum Mechanics and the Quantum Mechanics of Spacetime.}

\refis{Har6} J.B.Hartle, in, {\it Proceedings of
the Cornelius Lanczos International Centenary Confererence},
edited by J.D.Brown, M.T.Chu, D.C.Ellison and R.J.Plemmons
(SIAM, Philadelphia, 1994)
%``Quasiclassical Domains in a Quantum 
%Universe'', preprint gr-qc/9404017 (1994).

\refis{Har7} J.B.Hartle, ``Quantum Pasts and the Utility of 
Histories", preprint gr-qc/9712001.

\refis{HaB} T.A.Brun and J. B. Hartle, {\sl Phys.Rev.} {\bf E59},
6370 (1999).

\refis{HLM} J. B. Hartle, R. Laflamme and D. Marolf, 
\pr {\bf D51}, 7007 (1995).
 
\refis{IsL} C.Isham and N.Linden, \pr {\bf A55}, 4030 (1997).
%"Information entropy and the space of decoherence functions
%in generalized quantum theory" is published:
% in fact it is at quant-ph/9612035.

\refis{JoZ} E.Joos and H.D.Zeh, {\sl Z.Phys.} {\bf B59}, 223 (1985).

\refis{McE} J.N.McElwaine, {\sl Phys.Rev.} {\bf A53}, 2021 (1996).
%Approximate and Exact Consistency of Histories

\refis{Men} M.Mensky, {\sl Sov. Phys. JETP}, {\bf 50}, 667 (1979);
{\sl Theor. Math. Phys.} {\bf 75}, 357 (1988);
{\it Continuous Quantum Measurements and Path Integrals},
(IOP Publishing, Bristol and Philadelphia, 1993).

\refis{Omn} R. Omn\`es, {\sl J.Stat.Phys.} {\bf 53}, 893 (1988);
{\bf 53}, 933 (1988);
{\bf 53}, 957 (1988);
{\bf 57}, 357 (1989);
{\sl Ann.Phys.} {\bf 201}, 354 (1990);
{\sl Rev.Mod.Phys.} {\bf 64}, 339 (1992).

%\refis{Omn1} R.Omn\`es, \pr {\bf A56}, 3383 (1997).
% General theory of the decoherence effect in quantum mechanics.

\refis{QBM} G.S.Agarwal, \pr {\bf A3}, 828 (1971); \pr {\bf A4}, 739 (1971);
H.Dekker, \pr {\bf A16}, 2116 (1977); {\sl Phys.Rep.} {\bf 80}, 1 (1991);
G.W.Ford, M.Kac and P.Mazur, \jmp {\bf 6}, 504 (1965);
H.Grabert, P.Schramm, G-L. Ingold, {\sl Phys.Rep.} {\bf 168}, 115 (1988);
V.Hakim and V.Ambegaokar, \pr {\bf A32}, 423 (1985);
J.Schwinger, \jmp {\bf 2}, 407 (1961);
I.R.Senitzky, \pr {\bf 119}, 670 (1960).

%\refis{Qua} Steane. Quantum Computing.,

\refis{Whe} J.A.Wheeler, in, {\it Complexity, Entropy and the
Physics of  Information}, edited by W.Zurek (Addison--Wesley,
Redwood City, CA, 1990).

\refis{Woo} W.Wootters, ``The Aquisition of Information from
Quantum Measurements'', University of Texas PhD thesis (1980).

\refis{Zur1} See for example, J.P.Paz and W.H.Zurek, \pr {\bf D48},
2728 (1993); W.Zurek, in {\it Physical Origins of Time Asymmetry},
edited by  J.J.Halliwell, J.Perez-Mercader and W.Zurek (Cambridge
University Press, Cambridge, 1994). Zurek's contributions
to the subject are extensive and the above is only a
suggestive selection.

\refis{Zur2} W.Zurek, {\sl Phil.Trans.R.Soc.Lond.A} {\bf 356}, 1793
(1998).
%``Decoherence, Einselection, and the
%Existential Interpretation (The Rough Guide)'', preprint
%quant-ph/9805065.

\refis{Zur3} An interesting collection of article on many aspects of
the physics of information may be found in, {\it Complexity, Entropy
and the Physics of  Information}, edited by W.Zurek
(Addison--Wesley, Redwood City, CA, 1990).

\refis{Zur4} W.Zurek, \pr {\bf D24}, 1516 (1981).
% Pointer basis of quantum apparatus: into which
% mixture does the wavepacket collapse?

\refis{Zur5} W.Zurek, \pr {\bf D26}, 1861 (1982).
% Environment induced superselection rules.

\refis{Zur6} W.Zurek, in {\it Quantum Optics,
Experimental Gravitation and Measurement Theory}
(page 87), edited by P.Meystre and M.O.Scully
(NATO ASI Series, Plenum, New York, 1983).

\endreferences

\end